\documentclass[12pt,twoside]{article}

\usepackage[utf8]{inputenc}
\usepackage[english]{babel}
\usepackage[numbers,sort&compress]{natbib}
\usepackage[colorlinks=true, linkcolor=blue, bookmarks=true, linktocpage=true]{hyperref}
\usepackage{graphicx}
\usepackage{tabularx}
\usepackage[table]{xcolor}
\usepackage{tcolorbox}
\tcbuselibrary{theorems}
\usepackage{caption}
\usepackage{anysize}
\usepackage{subcaption}
\usepackage{amsmath}
\usepackage{amssymb}
\usepackage{tensor}
\usepackage{braket}
\usepackage{pdfpages}

\allowdisplaybreaks[1] %% Auto page breaking with long equations

\newcommand{\cC}{\mathcal{C}}

\newcommand{\cH}{\mathcal{H}}
\newcommand{\cM}{\mathcal{M}}

\newcommand{\cO}{\mathcal{O}}

\newcommand{\cQ}{\mathcal{Q}}

\newcommand{\bx}{\mathbf{x}}
\newcommand{\by}{\mathbf{y}}
\newcommand{\bz}{\mathbf{z}}
\newcommand{\bX}{\mathbf{X}}
\newcommand{\barA}{\bar{A}}
\newcommand{\rd}{\mathrm{d}}

\newcommand{\hg}{\hat{g}}

\newcommand{\hP}{\hat{P}}
\newcommand{\dx}{\left( \prod_{i=1}^L \prod_{a=1}^N \rd x^a_i \right)}

\marginsize{1.1in}{1.1in}{0.7in}{1.1in}
\numberwithin{equation}{section}

\begin{document}

\begin{titlepage}

\begin{center}

\phantom{ }
\vspace{1cm}

{\bf \Large{Definitions of entwinement}}
\vskip 0.7cm
Ben Craps,${}^{\text a}$ Marine De Clerck,${}^{\text b}$ Alejandro Vilar L\'opez${}^{\text c}$
\vskip 0.05in

\vspace{.6cm}

\small{${}^{\text{a}}$ \textit{Theoretische Natuurkunde, Vrije Universiteit Brussel (VUB) and\\
The International Solvay Institutes, Pleinlaan 2, B-1050 Brussels, Belgium}}

\vspace{.3cm}

\small{${}^{\text{b}}$ \textit{Department of Applied Mathematics and Theoretical Physics, \\
University of Cambridge, Cambridge CB3 0WA, United Kingdom}}

\vspace{.3cm}

\small{${}^{\text{c}}$ \textit{Physique Théorique et Mathématique and International Solvay Institutes, \\
Université Libre de Bruxelles (ULB); C.P. 231, 1050 Brussels, Belgium}}

\vspace{.8cm}
\texttt{\href{mailto:Ben.Craps@vub.be}{Ben.Craps@vub.be}, 
\href{mailto:md989@cam.ac.uk}{md989@cam.ac.uk}, 
\href{mailto:alejandro.vilar.lopez@ulb.be}{alejandro.vilar.lopez@ulb.be}}

\vspace{2cm}

\begin{abstract}
Entwinement was first introduced as the CFT dual to extremal, non-minimal geodesics of quotiented ${\rm AdS}_3$ spaces. It was heuristically meant to capture the entanglement of internal, gauged degrees of freedom, for instance in the symmetric product orbifold CFT of the D1/D5 brane system. %Following up on the original work, 
The literature now contains different, and sometimes inequivalent, field theory definitions of entwinement. In this paper, we build a discretized lattice model of symmetric product orbifold CFTs, and explicitly construct a gauge-invariant reduced density matrix whose von Neumann entropy agrees with the holographic computation of entwinement. Refining earlier notions, our construction gives meaning to the entwinement of an interval of given size within a long string of specific length. We discuss similarities and differences with previous definitions of entwinement.
\end{abstract}

\end{center}
\end{titlepage}

\tableofcontents

%%%%%%%%%%%%%%%%%%%%%%%%%%%%%%%%%%%%%%%%
\section{Introduction}
\label{sec:Intro}
%%%%%%%%%%%%%%%%%%%%%%%%%%%%%%%%%%%%%%%%

One of the most remarkable ideas that has arisen from holographic considerations in the last two decades is that spacetime geometry emerges from quantum entanglement. A concrete setup where one can explore this relation and its implications for gravity is the AdS/CFT correspondence. A cornerstone of the duality is the 
Ryu-Takayanagi (RT) prescription \cite{Ryu:2006bv,Ryu:2006ef},
\begin{equation}
\label{Intro:RyuTakayanagi}
S(A) = \underset{\gamma_A}{\rm{min}} \, \frac{{\rm Area}(\gamma_A)}{4 G_N} ~ ,
\end{equation}
which yielded the first concrete hint for a deep connection between the emergence of spacetime and the entanglement structure of the dual field theory, at least in the limit where the bulk is static and can be treated (semi-)classically.
The RT formula posits that the entanglement entropy of a constant-time boundary region $A$ corresponds to the area of a bulk surface that is anchored to the ends of $A$. The bulk surface of interest is the outcome of a minimization over bulk codimension-2 surfaces $\gamma_A$ subject to the following constraints: they are supported on the same constant-time slice as $A$, they end at the boundary $\partial A$, and they are homologous to $A$. This result is by now well-established to first order in a $G_N \sim 1/N$ expansion \cite{Lewkowycz:2013nqa}; generalizations to non-static situations \cite{Hubeny:2007xt} or higher order corrections are also available in the literature \cite{Faulkner:2013ana,Engelhardt:2014gca}.
Starting from the RT proposal, a whole new program developed whose aim was to investigate the intriguing interplay between the connectivity of classical geometry and the quantum entanglement between degrees of freedom in the dual field theoretic description \cite{VanRaamsdonk:2009ar,VanRaamsdonk:2010pw,Maldacena:2013xja}. Perhaps the most explicit evidence supporting the emergence of a smooth spacetime obeying the dynamical laws of classical gravity from entanglement can be attributed to \cite{Lashkari:2013koa,Faulkner:2013ica,Swingle:2014uza}, in which bulk equations of motion were perturbatively reproduced from the laws of quantum entanglement in the CFT. 

\begin{figure}[t!]
\centering
\begin{subfigure}{.40\textwidth}
  \centering
  \includegraphics[width=\linewidth]{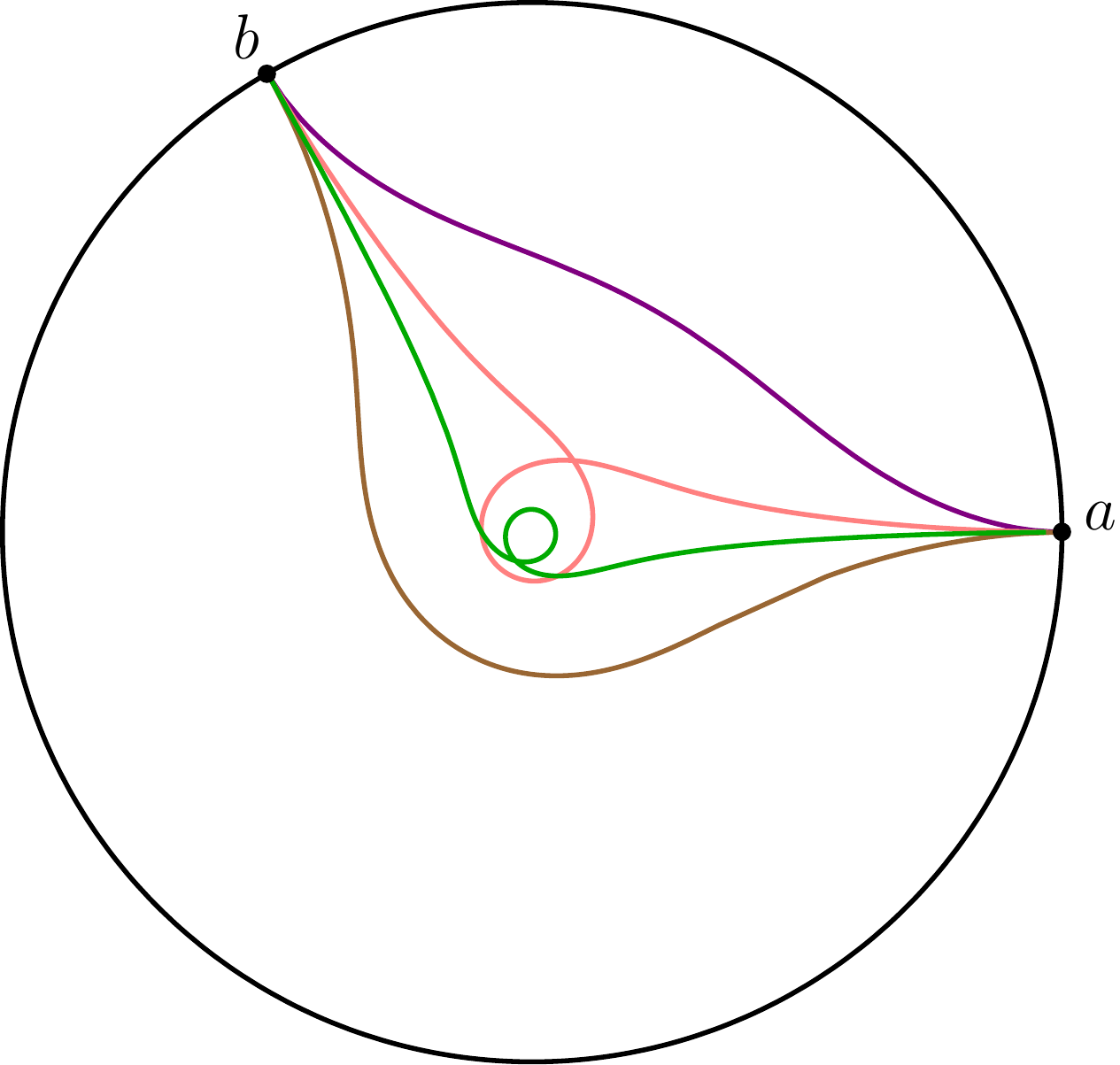}
\end{subfigure}
\begin{subfigure}{.40\textwidth}
  \centering
  \includegraphics[width=\linewidth]{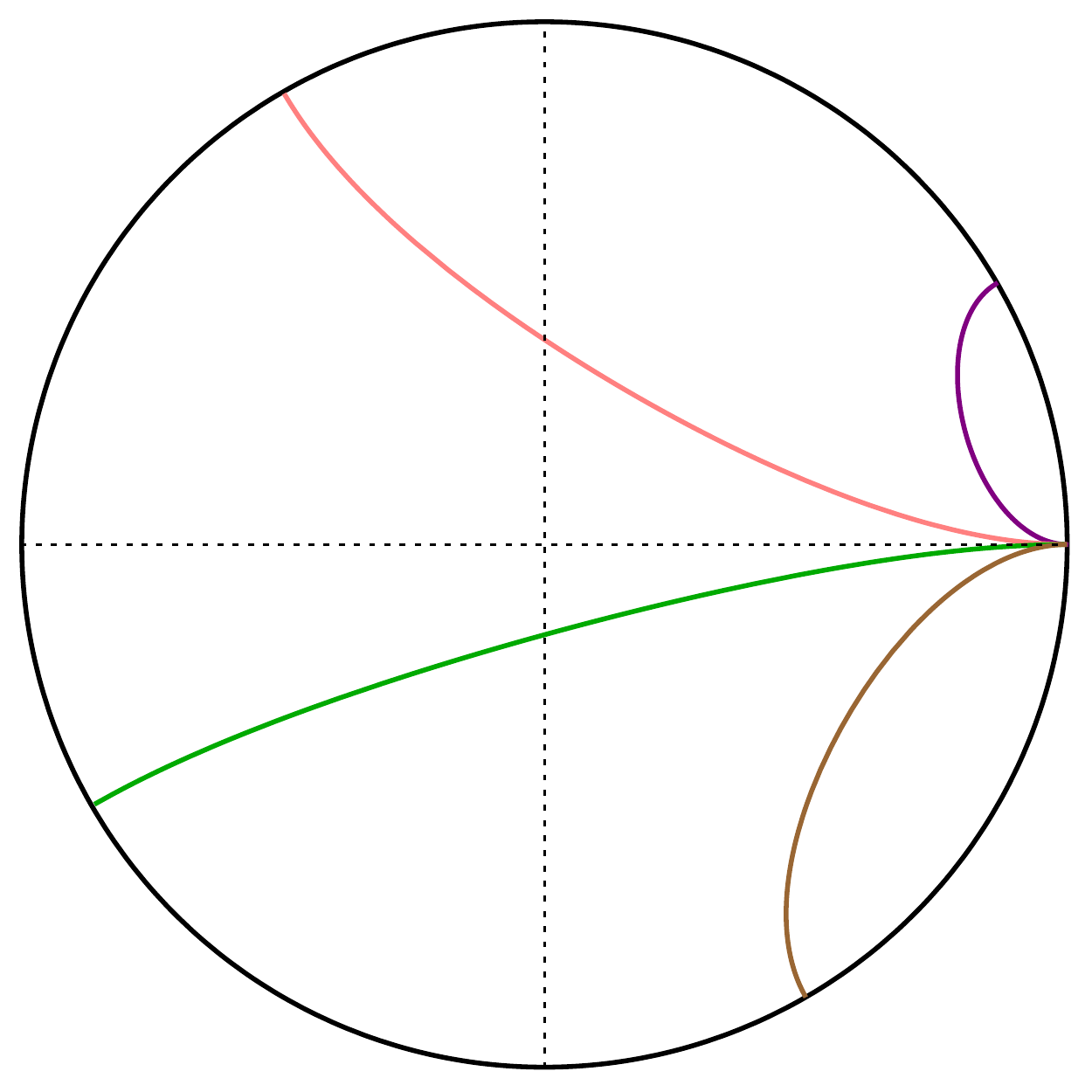}
\end{subfigure}
\caption{{\bf Left:} The $n$ geodesics between the boundary points $a$ and $b$, which delineate a region $A$, in the conical defect geometry AdS$_3/\mathbb{Z}_n$ with $n=4$. The minimal geodesic is the purple one. {\bf Right:} The same geodesics from the perspective of the covering space AdS$_3$.}
\label{fig: geodesics conical defect}
\end{figure}

Given a boundary region $A$, the RT prescription \eqref{Intro:RyuTakayanagi} instructs us to select the bulk surface anchored to $A$ that is the global minimum of the area functional, subject to the constraints described above. Yet, certain geometries allow for the existence of multiple area-extremizing surfaces anchored to a given boundary region, and it seems a natural and interesting question to ask if these extremal, non-minimal surfaces also have a dual representation in the CFT. A particularly convenient setup to try to answer this question is that of global $\rm{AdS}_3$ and the related quotient geometries ${\rm AdS}_3/\mathbb{Z}_n$. The latter spacetimes are constructed by identifying the global angular coordinate in $\rm{AdS}_3$ as $\phi \sim \phi + 2 \pi/n$, effectively creating a conical singularity at the origin $r=0$. One can picture constant-time surfaces of ${\rm AdS}_3/\mathbb{Z}_n$ as wedges of the full $\rm{AdS}_3$ disk with opening angle $2 \pi/n$. In a three-dimensional setting, co-dimension two surfaces are curves, and the geodesics of $\rm{AdS}_3$ are inherited by the quotient geometry upon imposing the $\mathbb{Z}_n$ identification. The equal-time geodesics of ${\rm AdS}_3/\mathbb{Z}_n$ wind around the conical defect whenever the original $\rm{AdS}_3$ geodesic covers an angular range wider than $ \pi/n$, as illustrated in Figure~\ref{fig: geodesics conical defect}.
For every connected boundary region $A$ of the conical defect geometry, one identifies $n$ geodesics anchored to $\partial A$. Notice that the minimal geodesics cannot penetrate arbitrarily deep into the bulk. This leaves a so-called \emph{entanglement shadow} \cite{Balasubramanian:2014sra,Freivogel:2014lja} in the middle of the conical defect spacetime: a bulk region whose local geometry appears inaccessible through the entanglement entropy of spatial subregions in the boundary theory. Extremal, non-minimal curves, on the other hand, penetrate deeper into the bulk than the minimal ones, reaching points arbitrarily close to the conical singularity.
In view of the program aiming at reconstructing geometry from the entanglement of the dual theory, this represented a substantial motivation to investigate whether these winding geodesics have an interesting counterpart in the boundary theory.\footnote{As a matter of fact, we know that any notion in the dual captured by extremal surfaces in the bulk is not going to be enough to reconstruct the full geometry in general spacetimes \cite{Engelhardt:2013tra}. It is still interesting, though, to understand the gravitational dual of non-minimal extremal surfaces, and there are cases like the one we described in which they do probe the bulk deeper than the minimal surfaces.} We mention in passing that entanglement shadows have a typical size comparable to the AdS scale, and reconstructing this region may also shed light on the long-standing conundrum of understanding sub-AdS locality in AdS/CFT \cite{Susskind:1998dq,Susskind:1998vk,Heemskerk:2009pn}. In this context, as well as in models without spatial extent such as matrix models \cite{Banks:1996vh,Kazakov:2000pm}, it is expected that the entanglement between internal rather than spatial degrees of freedom is an essential ingredient of the duality. One may indeed hope that general lessons can be learned from understanding the entanglement shadows of these simple spacetimes in detail.

The CFT notion that encodes the length of non-minimal extremal curves in a conical defect was introduced in \cite{Balasubramanian:2014sra} under the name of \emph{entwinement}. Generalizing conventional entanglement entropy, entwinement was proposed as a measure for the entanglement between internal degrees of freedom that are not necessarily spatially organized, as expected from our discussion above. Concretely, this quantity was initially computed by considering the covering bulk and boundary theories obtained by ungauging the $\mathbb{Z}_n$ discrete gauge symmetry of the AdS$_3$ quotient. Since non-minimal geodesics in the conical defect spacetime descend from minimal geodesics on the covering space, their length is dual to standard, spatial entanglement entropy on the covering boundary theory. After symmetrizing the result over $n$ translated embeddings of the region $A$ homologous to the geodesic in the covering space, to produce a $\mathbb{Z}_n$ invariant result, one obtains entwinement. 
This procedure is interpreted as effectively quantifying the entanglement between non-spatially organized degrees of freedom. Indeed, from the viewpoint of the $\mathbb{Z}_n$-symmetric theory, it departs from the usual picture in which one computes entanglement entropy between a certain spatial region and its complement. However, this approach has an evident drawback: though the result is gauge-invariant, the process requires going to the covering theory, and thus introduces unphysical states.

A gauge-invariant alternative to the original definition of entwinement, based on the replica trick, was proposed in \cite{Balasubramanian:2016xho}. This approach formulates entwinement for the first time as an entanglement measure for a precise set of degrees of freedom in the boundary CFT state dual to the conical defect spacetime (as well as other quotients of AdS$_3$). Let us briefly recall the content of the boundary theory describing these spacetimes.
The dual description of three-dimensional, asymptotically AdS spacetimes is obtained by considering a bound state of D1 and D5 branes in type IIB string theory (see \cite{Giveon:1998ns} for details). 
Weakly coupled strings propagating on ${\rm AdS}_3 \times S^3 \times T^4$ are found to be dual to a 2d (S)CFT whose moduli space contains a special point, called the orbifold point, for which the dynamics becomes a free sigma model with target space $(T^4)^N/S_N$. The parameter $N$, which corresponds to the product of the number of D1 and D5 branes in the string theory setup, is taken to be large and tunes the ratio of the ${\rm AdS}_3$ radius to the three-dimensional Planck length $G_N$. The theory at the orbifold point is a symmetric product orbifold CFT and contains $N$ copies of four bosons (and their supersymmetric partners) defined on a circle. They parameterize the embedding of closed string configurations in the $N$ copies of $T^4$. The $S_N$ quotient is understood as a discrete gauge symmetry on the $N$ copies and allows for the existence of \textit{twisted sectors},\footnote{From the CFT point of view, these additional sectors are required to maintain modular invariance \cite{Dixon:1985jw,Ginsparg:1988ui}. In this paper, we will argue for them from the perspective of a discretized lattice gauge theory.} in which the fields are periodic up to $S_N$ transformations when going around the spatial circle of the CFT. 

The field theory dual to a conical defect geometry can be 
constructed by applying a particular twist operator \cite{Balasubramanian:2000rt,Martinec:2001cf,Martinec:2002xq,Balasubramanian:2005qu}, which partitions the $N$ pieces of strings into $N/n$ long strings of length $n$ in the target space. We note that the free orbifold point of the CFT is far in moduli space from those points which have a semiclassical gravitational description, for which the field theory would be strongly coupled. Still, it is a general expectation that certain quantities (e.g., those which are protected by supersymmetry or can be obtained by means of the covering theory \cite{Martinec:2002xq}) can be reliably computed at the free orbifold point. Using this description, the replica trick approach of \cite{Balasubramanian:2016xho} precisely identifies the CFT internal gauged degrees of freedom whose entanglement entropy corresponds to entwinement, in a manifestly gauge-invariant manner.

Perhaps the most standard approach to conventional entanglement entropy is, however, based on the reduced density matrix on a boundary region $A$, whose von Neumann entropy provides a definition for the entanglement entropy of the degrees of freedom in $A$. This avenue to entanglement entropy naturally fits in the framework of algebraic QFT, where the central objects are (sub)algebras of operators. In theories where the Hilbert space $\cH$ factorizes, $\cH = \cH_A\otimes \cH_{\bar{A}}$, there exists a natural subalgebra of operators $\mathcal{A}$ associated to the factor $\cH_A$ (we usually think of $A$ as a spatial subregion). Given a state $\rho$ in $\cH$, this subalgebra contains a unique element $\rho_\mathcal{A}$ such that
\begin{equation}
{\rm Tr}_{\mathcal{H}_A}[\rho_A\mathcal{O}] = \langle \mathcal{O} \otimes \mathbf{1}_{\bar{A}} \rangle_{\rho}  
\label{intro: reduced density matrix}
\end{equation}
for every $\mathcal{O} \in \mathcal{A}$. The entanglement entropy of $A$ is then given by the von Neumann entropy of this reduced density matrix, 
\begin{equation}
S_{EE} = - {\rm Tr}_{\mathcal{H}_A}[\rho_A \log \rho_A] ~ ,
\label{intro: conventional EE}
\end{equation}
 and is a measure for the amount of information hidden from measurements on that subsystem alone. This approach, however, is complicated by the presence of a gauge symmetry, since the resulting gauge constraints interfere with the factorization property of the Hilbert space. For instance, the definition of entanglement entropy for a spatial bi-partition is notably subtle in the presence of gauge symmetries \cite{Buividovich:2008gq,Buividovich:2008yv,Donnelly:2011hn,Casini:2013rba,Radicevic:2014kqa,Donnelly:2014gva,Ghosh:2015iwa,Aoki:2015bsa,Soni:2015yga,VanAcoleyen:2015ccp}. In these contexts, the tensor factorization of the Hilbert space is impeded by the existence of non-local physical degrees of freedom such as Wilson loops, which do not belong to either of the complementary subregions in the spatial bi-partition. In the symmetric product orbifold setup, the $S_N$ gauge symmetry permutes internal degrees of freedom, which puts constraints on the physical states of the theory. Hence,  the resulting physical Hilbert space does not admit a tensor factor decomposition for bi-partitions that assign degrees of freedom living at the same spatial point to two different subsets. This picture is analogous to studies of entanglement entropy in systems of identical particles \cite{schliemann2001double,schliemann2001quantum,pavskauskas2001quantum,eckert2002quantum,sciara2017universality}.
 
The problem of formally defining entwinement was approached from this algebraic perspective in the works \cite{Balasubramanian:2018ajb,Lin:2016fqk,Erdmenger:2019lzr}. The goal of \cite{Balasubramanian:2018ajb} was to define a gauge-invariant reduced density matrix for a general bi-partition of the internal degrees of freedom of a symmetric product orbifold theory. This algebraic approach was found to agree with previous definitions of entwinement when one considers the entanglement of a single strand in the D1/D5 orbifold CFT with its complement, but resulted in diverging results otherwise. This disagreement was interpreted in their conclusions in the following terms. While entwinement was previously understood as being a measure for the entanglement of \emph{connected} strands, the proposed formalism of \cite{Balasubramanian:2018ajb} produces unwanted additional disconnected pieces which also contribute to the entanglement entropy. A gauge-invariant density matrix relevant to entwinement would hence need to deal with this connectedness in an appropriate way. 
A second output of their analysis was the realization that, in contrast to conventional (spatial) reduced density matrices, the reduced density matrix associated to the internal gauged degrees of freedom relevant for entwinement is instead associated to a linear subspace of operators \cite{Balasubramanian:2018ajb} (see also \cite{Erdmenger:2019lzr}). This finding has been somewhat in tension with \cite{Lin:2016fqk}, which focuses on defining entwinement as the algebraic entanglement entropy of a gauge-invariant reduced density matrix that is associated with an actual subalgebra of gauge-invariant operators. An interesting construction for this reduced density matrix and subalgebra was detailed in the context of a toy model for the D1/D5 orbifold CFT \cite{Lin:2016fqk}, and was argued to hold more generally.

The original proposal \cite{Balasubramanian:2014sra} has also been recast in algebraic terms in \cite{Erdmenger:2019lzr} via the extended Hilbert space method. This language was used to show the equivalence between the original approach and the interpretation of entwinement as the minimal entropy of the probability distribution for density matrices resulting from measurements constrained to the linear subspace of operators associated with non-spatial subsets of degrees of freedom. This was shown for general CFTs with $\mathbb{Z}_n$ gauge symmetry.

In the present paper, our goal is to revisit the approach of \cite{Balasubramanian:2018ajb} by constructing a gauge-invariant reduced density matrix for entwinement from the point of view of lattice gauge theories. Our purpose is to clarify how the notion of connectedness that was missing in \cite{Balasubramanian:2018ajb} can be implemented naturally by properly taking into account the gauge transformations of all the degrees of freedom.\footnote{Indeed, it was remarked in section~6 of \cite{Erdmenger:2019lzr} that the action of symmetrization on a state representing a connected target space configuration should normally not produce the disconnected pieces found in \cite{Balasubramanian:2018ajb}.}  As a consequence, our conclusion will be different from the one presented in \cite{Balasubramanian:2018ajb}: there is indeed a linear subspace of operators in the symmetric product orbifold theory for which an associated reduced density matrix computes entwinement via its von Neumann entropy. While the specific type of bi-partition of the degrees of freedom that gives rise to entwinement will be the main focus of this work, we remark that our formalism is applicable to more general situations. We will therefore allow for an arbitrary subset of degrees of freedom and twisted sector throughout the paper, and only commit to the specific cases relevant to the holographic setup when setting the stage for comparing with other approaches. In contrast to some of the other methods, we shall quantify entanglement between internal degrees of freedom in theories with a discrete gauge symmetry, without essentially relying on the introduction of unphysical states.

The paper is organized as follows. In section~\ref{sec:Orbifolds as discrete gauged theories} we formulate a discretized model for symmetric product orbifold theories in terms analogous to those of standard lattice gauge theories. We emphasize the role of the link variables in the emergence of twisted sectors in the Hilbert space.
In section~\ref{sec:EntwinementAsEntanglement} we define gauge-invariant reduced density matrices that reproduce entwinement via their von Neumann entropy. We identify the linear subspaces of operators associated with these reduced density matrices in section~\ref{subsec:ReducedDensityMatrixEntwinement}, and we make contact with earlier definitions of entwinement in  section~\ref{subsec:DefinitionsOfEntwinement} by focusing on the specific case of the D1/D5 orbifold CFT. We conclude with possible future directions in section~\ref{sec:Discussion}.
Appendix~\ref{App:Algebraic entwinement in $S_3$-symmetric spin model} contains a technical note about the construction proposed in \cite{Lin:2016fqk} to obtain a reduced density matrix for entwinement associated to a subalgebra of gauge-invariant operators instead of a linear subspace. We find it impossible to extend the proposal beyond the simplified $\mathbb{Z}_2$ setup of \cite{Lin:2016fqk}, and discuss in detail the next simplest example which provides a counterexample to the general validity of the method.

%%%%%%%%%%%%%%%%%%%%%%%%%%%%%%%%%%%%%%%%
\section{Symmetric product orbifolds as lattice gauge theories}
\label{sec:Orbifolds as discrete gauged theories}
%%%%%%%%%%%%%%%%%%%%%%%%%%%%%%%%%%%%%%%%

In this section, we introduce the Hilbert space content of symmetric product orbifold theories with target space $M^N/S_N$ from the point of view of lattice gauge theories. This is not the usual setup for orbifold 2d CFTs. However, essential aspects for our purposes are most explicit in the lattice model and it is worthwhile to start on the lattice and take the continuum limit at the end. The connections with the more conventional setup will become clear as we proceed. For concreteness, we will restrict our discussion to symmetric product orbifold theories, with discrete $S_N$ symmetry, though the generalization to other discrete symmetry groups is straightforward. 
This section is meant to set the notation and conventions for section~\ref{sec:EntwinementAsEntanglement}, where we shall revisit the algebraic characterization of entwinement of \cite{Balasubramanian:2018ajb} in terms of gauge-invariant reduced density matrices. 

\subsection{Constructing the lattice theory}

Let us start by considering a \textit{seed} theory, which consists of a 1+1 dimensional lattice model of bosonic dynamical variables $X$. We implicitly allow for multi-component sets, $X \equiv (X^{(\mu)}) = (X^{(0)}, X^{(1)}, \dots, X^{(D-1)})$, which describe the embedding of the lattice points in a $D$-dimensional manifold $M$. Borrowing the nomenclature from string theory models, we will refer to these variables as \emph{target space coordinates}, although most of the time the index $\mu$ will not play any significant role and we will suppress it to avoid clutter. These dynamical variables are defined on a discretization of the circle, and we label the variables at different points as $X_i$, $i = 1, \dots, L$, with the identification $X_{L+1} = X_1$. The detailed dynamics of this seed theory is not going to be essential, but to have something definite in mind we shall consider the Hamiltonian\footnote{We assume a Euclidean metric for the target space coordinates whenever we use the compact notation $(X_{i+1} - X_i)^2$.}
\begin{equation}
\label{Discrete:SeedHamiltonian}
H_{\rm seed} = \sum_{i=1}^L \frac{\delta}{2} \left[ \Pi_i^2 + \frac{1}{\delta^2} (X_{i+1} - X_i)^2 + \dots \right] ~ ,
\end{equation}
where $\delta$ is the lattice spacing, $\Pi_i$ the momentum conjugate to $X_i$, and the dots represent local interactions.

We now consider $N$ copies of this seed theory. We label\footnote{We emphasize that the Latin label indicating the copy should not be confused with the Greek label defined above, which stands for the dimensions within the target space manifold $M$. The reader can assume that any upper index for $X$ used in the remainder of this paper denotes one of the $N$ copies of the symmetric product orbifold.} the dynamical variables $X^a_i$, with $a = 1, \dots, N$; and we consider Hilbert space operators $\hg_i$ which implement permutations at a single lattice point $i$. For a permutation $g_i \in S_N$, the action of the corresponding permutation operator $\hg_i$ on the fields is
\begin{equation}
\label{Discrete:SNActionOnLatticePoints}
\hg_i X^a_i \hg^{-1}_i = X^{g_i(a)}_i ~ , \qquad \hg_i \Pi^a_i \hg^{-1}_i = \Pi^{g_i(a)}_i ~ .
\end{equation}
It will be convenient in the following to adopt a matrix-like notation for these transformations. Thus, we pack all the $N$ copies of the variables at each lattice point into a (column) vector, which we denote $\bX_i$ or $\boldsymbol{\Pi}_i$, and write the previous transformations as
\begin{equation}
\label{Discrete:SNActionOnLatticePointsMatrix}
\hg_i \bX_i \hg^{-1}_i = g_i^{-1} \bX_i ~ , \qquad \hg_i \boldsymbol{\Pi}_i \hg^{-1}_i = g_i^{-1} \boldsymbol{\Pi}_i ~ ,
\end{equation}
where now $g_i$ is the $N \times N$ matrix representation implementing the $S_N$ transformation.\footnote{As an example, if we consider $S_3$, the matrix representation of $(1 2 3)$ is 
$(1 2 3) = \begin{pmatrix}
0 & 0 & 1 \\
1 & 0 & 0 \\
0 & 1 & 0
\end{pmatrix}$.
It is necessary to consider the inverse of $g_i$ in \eqref{Discrete:SNActionOnLatticePointsMatrix} to reproduce \eqref{Discrete:SNActionOnLatticePoints}, and also to provide a valid representation via left-multiplication by the corresponding matrices.}

Summing $N$ copies of the seed Hamiltonian \eqref{Discrete:SeedHamiltonian}, it is immediate to see that terms defined at a single lattice point are invariant under permutations:
\begin{equation}
\label{Discrete:InvarianceLocalInteractions}
 \hg_i \left( \sum_{a=1}^N X_i^a X_i^a \right) \hg_i^{-1} = \sum_{a=1}^N X^{g_i(a)}_i X^{g_i(a)}_i = \sum_{a=1}^N X_i^a X_i^a ~ .
\end{equation}
This argument continues to apply for terms with $\Pi_i^a$, or higher order terms in $X_i^a$. However, the nearest-neighbor couplings -- coming from spatial derivative terms in the continuum -- manifestly spoil the invariance of the $N$-copied Hamiltonian under local $S_N$ permutations. This can be remedied by the addition of a background (non-dynamical) \emph{gauge} field $U_{i+1,i}$ to the nearest-neighbor interactions, such that the total Hamiltonian takes the form
\begin{equation}
\label{Discrete:HamiltonianSymmetric}
H_{S_N} = \sum_{i=1}^L \frac{\delta}{2} \left[ \boldsymbol{\Pi}^T_i \boldsymbol{\Pi}_i + \frac{1}{\delta^2} (\bX_{i+1} - U_{i+1,i} \bX_i)^T (\bX_{i+1} - U_{i+1,i} \bX_i) + \dots \right] ~ .
\end{equation}
The background gauge field lives on the links between lattice points and takes values in the group of $N \times N$ permutation matrices. The action of the local $S_N$ transformation on the gauge field is
\begin{equation}
\label{Discrete:SNActionOnLatticeLinks}
\hg_{i+1} U_{i+1,i} \hg_{i+1}^{-1} = g_{i+1}^{-1} U_{i+1,i} ~ , \qquad \hg_{i} U_{i+1,i} \hg_{i}^{-1} = U_{i+1,i} g_i ~ .
\end{equation}
Since the permutation matrices are orthogonal, $H_{S_N}$ is invariant under any local $S_N$ transformations implemented by the operators $\hg_i$.

The Hilbert space of our theory -- which we dub the \emph{covering theory}\footnote{The covering theory in the symmetric product orbifold theory should not be confused with the covering theory obtained by ungauging the $\mathbb{Z}_n$ symmetry of the quotient spacetime ${\rm AdS}_3/\mathbb{Z}_n$. In the following, the meaning of the term should be clear from the context.} -- is spanned by a basis of eigenstates of $\bX_i$ and $U_{i+1,i}$. We denote them by $\ket{\{ \bx_j \}, \{ u_{j+1,j} \} }$, reflecting the eigenvalues of the said operators:
\begin{subequations}
\label{Discrete:BasisStatesEigenvalues}
\begin{align}
\bX_i \ket{\{ \bx_j \}, \{ u_{j+1,j} \}} & = \bx_i \ket{\{ \bx_j \}, \{ u_{j+1,j} \}} ~ , \\
U_{i+1,i} \ket{\{ \bx_j \}, \{ u_{j+1,j} \}} & = u_{i+1,i} \ket{\{ \bx_j \}, \{ u_{j+1,j} \}} ~ .
\end{align}
\end{subequations}
We emphasize that, for every lattice point $i$, $\bx_i$ takes values in $M^N$, with $M$ the target space, and $u_{i+1,i}$ takes values in the standard matrix representation of $S_N$. The action of the local $S_N$ transformations on the basis states can be computed from their  effect on the corresponding operators,
\begin{equation}
\label{Discrete:SNActionOnStates}
\hg_i \ket{\{ \bx_j \}, \{ u_{j+1,j} \}} = \ket{\{ g_i \bx_i, \bx_j \}_{j \neq i}, \{u_{i+1,i} g_i^{-1}, g_i u_{i,i-1}, u_{j+1,j} \}_{j \neq i, i-1}} ~ .
\end{equation}
We take the basis states to be normalized as
\begin{equation}
\label{Discrete:NormalizationGeneralStates}
\braket{\{\bx_i \}, \{u_{i+1,i} \} | \{\by_i \}, \{v_{i+1,i} \}} = \prod_{i=1}^L \delta^{(N)} \left(\bx_i - \by_i \right) \delta_{u_{i+1,i}, v_{i+1,i}} ~ ,
\end{equation}
where we use a discrete Kronecker delta for the link variables since the $u_{i+1,i}$ take values in a finite set. It is immediate to verify that the $\hg_i$ are unitary operators with respect to this inner product. 

The form of the nearest-neighbor interactions in the Hamiltonian \eqref{Discrete:HamiltonianSymmetric} imposes some restrictions on the set of allowed states, since physical states should have finite energy in a continuum limit, $\delta \to 0$. These restrictions are the discrete manifestation of the continuity of field configurations in the continuum theory. Intuitively, the difference in field values between lattice points which are spatially close should not be too large. From the Hamiltonian, it is clear that one should impose
\begin{equation}
\label{Discrete:ContinuityConditionGeneral}
\bx_{i+1} - u_{i+1,i} \bx_i = \cO (\delta) \quad \forall i = 1, \dots, L
\end{equation}
on the eigenvalues of physical states. We call states satisfying this continuity condition \emph{allowed states}. The non-allowed states have infinite energy and effectively decouple from the allowed states in the Hilbert space. Notice that the transformation \eqref{Discrete:SNActionOnStates} implies that arbitrary local permutations map allowed states to allowed states. In the target space $M^N$, the corresponding states describe the embedding of the $N$ sets of $L$ discrete points on which the discrete continuity condition \eqref{Discrete:ContinuityConditionGeneral} is imposed.

Although the formalism developed above occasionally refers to the Hamiltonian \eqref{Discrete:HamiltonianSymmetric} for inspiration, we stress that the structure we are interested in is not tied to it. The essential ingredient for the next sections is that the Hilbert space of our $N$-copied product theory is spanned by the states $\ket{\{ \bx_j \}, \{ u_{j+1,j} \} }$, which describe the field configuration of bosonic variables on which $S_N$ permutations act according to \eqref{Discrete:SNActionOnLatticePointsMatrix} and \eqref{Discrete:SNActionOnLatticeLinks}. Moreover, a discrete notion of continuity is implemented on states following \eqref{Discrete:ContinuityConditionGeneral}. Our goal is now to mod out the $S_N$ symmetry and interpret it as a gauge redundancy.

%%%%%%%%%%%%%%%%%%%%%%%%%%%%%%%%%%%%%%%%
\subsection{Gauging the $S_N$ symmetry}
%%%%%%%%%%%%%%%%%%%%%%%%%%%%%%%%%%%%%%%%

Up to now, the local $S_N$ symmetry has been a true symmetry of the covering theory: the transformation \eqref{Discrete:SNActionOnStates} maps states in the Hilbert space to different states which have the same energy. We can gauge the $\bigotimes_{i=1}^{L}S^{(i)}_N$ symmetry by projecting the states onto the $S_N$-invariant subspace by means of the operator
\begin{equation}
\label{Discrete:GaugeInvariantProjector}
\hP = \bigotimes_{i=1}^{L} \left( \frac{1}{N!} \sum_{g_i \in S_N} \hg_i \right) ~ .
\end{equation}
The normalization is chosen so that the operator defines a projection,
\begin{equation}
\label{Discrete:ProjectorSquaresToOne}
\hP^2 = \bigotimes_{i=1}^{L} \left( \frac{1}{(N!)^2} \sum_{g_i, h_i \in S_N} \hat{h}_i \hat{g}_i \right) = \bigotimes_{i=1}^{L} \left( \frac{1}{(N!)^2} \sum_{h_i, l_i \in S_N} \hat{l}_i \right) = \hP  ~ ,
\end{equation}
where in the second step we sum over $l_i = h_i g_i$ instead of $g_i$. As a consequence of the unitarity of the representation, $\hg_i^{\dagger} = \hg_i^{-1}$, it immediately follows that the projector is Hermitian, $\hP^{\dagger} = \hP$.

Let $\cH = \otimes_{a=1}^N \cH^{(a)}_{\rm seed}$ be the original $N$-copied product Hilbert space, and call $\cH_S = \hP \cH$ the space of completely symmetric states. The local $S_N$ transformations act trivially on the symmetric states, thus turning the symmetry into a gauge redundancy. How can we characterize these symmetric states? As a first step, we will simplify the description by performing a partial projection. The total symmetry group can be decomposed as a semidirect product $\bigotimes_{i=1}^{L}S^{(i)}_N = S_N \ltimes \left(\bigotimes_{i=1}^{L-1}S^{(i)}_N \right)$, where the action of a global $S_N$ subgroup has been singled out,
\begin{equation}
\label{Discrete:GlobalProjectorIsolation}
\hP = \left[ \frac{1}{N!} \sum_{g \in S_N} \hg \otimes \ldots \otimes \hg \right] \left[ \bigotimes_{i=1}^{L-1} \left( \frac{1}{N!} \sum_{g_i \in S_N} \hat{g}_i \right) \right] \equiv \hP_{gl} \hP_{L-1} ~ ,
\end{equation}
and where the second factor defines a projector $\hP_{L-1}$ which only includes group elements acting on the first $L-1$ lattice sites. Then, projecting the states in $\cH$ with $\hP_{L-1}$ defines a partially invariant space $\cH_{gl} = \hP_{L-1} \cH$, where we can introduce a set of convenient representative states. Indeed, since $\hP_{L-1} \hg_i = \hP_{L-1}$ for $i = 1, \dots, L-1$, any state in $\cH$ with non-trivial link variables $v_{i+1,i}$ on the $L-1$ first lattice sites will be identified after projection with a state with trivial link variables on these lattice sites. Specifically, for any vertex variables $\{\by_i\}$ and link variables $\{ v_{i+1, i} \}$, one can write
\begin{equation}
\label{Discrete:RepresentativeStatesDerivation}
\hP_{L-1} \ket{\{ \by_i \}, \{ v_{i+1, i} \}} = \hP_{L-1} \ket{\{ \bx_i \}, \{ 1, \dots, 1, u_{1, L} \}} ~ ,
\end{equation}
where the variables on the right relate to those on the left as $\bx_i = g_i \by_i$, $u_{1,L} = g_1 v_{1, L}$, with
\begin{align}
\label{Discrete:TransformationsToRepresentative}
\nonumber g_{L} & = 1 ~ , \\
\nonumber g_{L-1} & = v_{L,L-1} ~ , \\
\nonumber g_{L-2} & = g_{L-1} v_{L-1,L-2} = v_{L,L-1} v_{L-1,L-2} ~ , \\
\nonumber \vdots & \\
g_1 & = g_2 v_{2,1} = v_{L,L-1} v_{L-1,L-2} \dots v_{2,1} ~ .
\end{align}
The outcome of this procedure is that we have sequentially \emph{gauged away} the link variables at every site, except for the last one. This can be viewed as a gauge-fixing procedure: we choose a convenient representative state among all those which are related by the transformations we want to gauge, and we use it to label the invariant states in the gauge (i.e., projected) theory. All states in the partially invariant space $\cH_{gl}$ are then characterized by the lattice point variables, $\bx_i$, as well as an $S_N$ element that sits at the link joining the last and first lattice sites. Choosing a convenient normalization, we will write them as
\begin{equation}
\label{Discrete:StatesInGlobalSpace}
\ket{ \{ \bx_i \}, u}_{gl} \equiv \sqrt{(N!)^{L-1}} \hP_{L-1} \ket{\{ \bx_i \}, \{ 1, \dots, 1, u \}} ~ .
\end{equation}
These states, which act as a generating set for $\cH_{gl}$, inherit the inner product
\begin{equation}
\label{Discrete:InnerProductGlobalSpace}
\sideset{_{gl}}{_{gl}}{\mathop{\braket{ \{ \by_i \} , v | \{ \bx_i \}, u }}} = \delta_{u,v} \prod_{i=1}^L \delta^{(N)} \left(\bx_i - \by_i \right) ~ ,
\end{equation}
where we used \eqref{Discrete:NormalizationGeneralStates},
\eqref{Discrete:GlobalProjectorIsolation}, \eqref{Discrete:StatesInGlobalSpace},  and the fact that $\hP^\dagger_{L-1}\hP_{L-1}=\hP_{L-1}^2=\hP_{L-1}$. The continuity condition \eqref{Discrete:ContinuityConditionGeneral} now constrains the lattice point variables in the  representative state \eqref{Discrete:StatesInGlobalSpace} to satisfy
\begin{equation}
\label{Discrete:ContinuityConditionGlobal}
\bx_{i+1} - \bx_i = \cO(\delta) \quad \forall i \neq L ~, \qquad \bx_1 - u \bx_L = \cO(\delta) ~ .
\end{equation}
In the target space, the representatives of the states \eqref{Discrete:StatesInGlobalSpace} describe the discretized version of $N$ continuous \textit{strands}, i.e., the embedding of a string segment in a higher dimensional manifold, with $u$ instructing how to glue the ends of the strands into \emph{long strings} (sets of strands forming a closed loop).

Notice, however, that since we have not yet gauged global transformations in $\cH_{gl}$, a residual global $S_N$ symmetry remains. These residual symmetries act on partially projected states as
\begin{equation}
\label{Discrete:ResidualGlobalSymmetry}
%\hg_{g} \ket{ \{ \bx_i \}, u}_g =
\left( \hg \otimes \dots \otimes \hg \right) \ket{ \{ \bx_i \}, u}_{gl} = \ket{ \{ g \bx_i \}, g u g^{-1}}_{gl} ~ ,
\end{equation}
as can be easily checked by moving the $\hg$ operators to the right of $\hP_{L-1}$. Since the fully gauge-invariant Hilbert space $\cH_S$ can be obtained by implementing global $S_N$ symmetrization on $\cH_{gl}$ with $\hP_{gl}$, gauge-invariant states can be defined as
\begin{equation}
\label{Discrete:GlobalSymmetryzation}
\ket{ \{ \bx_i \}, u }_S \equiv \sqrt{N!} \hP_{gl} \ket{ \{ \bx_i \}, u}_{gl} = \frac{1}{\sqrt{N!}} \sum_{g \in S_N} \ket{ \{ g \bx_i \}, g u g^{-1}}_{gl}  ~ .
\end{equation}
Note that, as emphasized in a related context in \cite{Erdmenger:2019lzr}, the action of a group element $\hg$ transforms both the vertex \textit{and} the link variable in such a way that different terms in \eqref{Discrete:GlobalSymmetryzation} (or even \eqref{Discrete:StatesInGlobalSpace}) describe the same target space configuration, albeit using a different labeling. In particular, the continuity constraints are invariant under permutations from the perspective of the target space, only the specifics of which variable \textit{on the lattice} produces a given target space point changes.

There is a redundancy in the definition of the symmetric states \eqref{Discrete:GlobalSymmetryzation}, since
\begin{equation}
\label{Discrete:RedundancyGlobalTransformation}
\ket{ \{ \bx_i \}, u }_S = \ket{ \{ g \bx_i \}, g u g^{-1}}_S
\end{equation}
for any $g \in S_N$. Thus, in the fully symmetrized space, the relevant link variable is only defined up to conjugation by any group element. Just as in \eqref{Discrete:StatesInGlobalSpace}, we fix this remaining redundancy by gauge fixing: we pick a representative in the symmetrized superposition of states to label the state. In this case, this amounts to choosing a representative element in each conjugacy class of $S_N$, and assigning it to the link variable labeling the state, which unambiguously defines the continuity of the vertex variables according to \eqref{Discrete:ContinuityConditionGlobal}. States in the \emph{orbifold theory} with Hilbert space $\cH_S$ are therefore spanned by 
\begin{equation}
\label{Wavefunction:BasisGaugeInvariantSubspace}
\cH_S = \text{span}\left\{ \ket{ \{\bx_i\}, u }_S \, | \, \bx_i \in \cM^N, u \in \cC(S_N) \right\} ~ ,
\end{equation}
with $\cC(S_N)$ a set containing one representative element of each conjugacy class. Thus, from our lattice construction, we recover the well-known fact that the Hilbert space of a symmetric product orbifold theory splits into a direct sum of twisted sectors, labeled by the conjugacy classes of $S_N$. 

It is important to note that this set of states in \eqref{Wavefunction:BasisGaugeInvariantSubspace} is not quite yet a basis for $\cH_S$, because there is still some redundancy left in the labeling of the vertex variables $\bx_i$. This is most easily seen when considering the following decomposition of the $S_N$-sum in \eqref{Discrete:GlobalSymmetryzation}. Given $\ket{ \{ \bx_i \}, u}_{S}$, consider the centralizer of $u$, denoted by $C_u$, which consists of all the elements in $S_N$ commuting with $u$. Since the centralizer defines a subgroup of the original symmetry group, we can decompose $S_N$ into left cosets. Selecting a representative from each coset we form a transversal set $K = \{ k_1, \dots, k_Q\}$, with $Q = N!/|C_u|$, so that any element in $S_N$ can be written as $g = k h$ with $k \in K$ and $h \in C_u$. Since $h u h^{-1} = u$, this gives a decomposition of the sum in \eqref{Discrete:GlobalSymmetryzation} as
\begin{equation}
\label{Discrete:GlobalSymmetryzationCosets}
\ket{ \{ \bx_i \}, u }_S = \sqrt{\frac{|C_u|}{N!}} \sum_{k \in K} \frac{1}{\sqrt{|C_u|}} \sum_{h \in C_u} \ket{ \{ k h \bx_i \}, k u k^{-1}}_{gl} ~ .
\end{equation}
We see that the sum over $K$ moves us between different link variables within the same conjugacy class, while, at fixed $k$, the sum over $C_u$ just reshuffles the vertex variables. 
Clearly, then,
\begin{equation}
\label{Discrete:VertexRedundancy}
\ket{ \{ \bx_i \}, u }_S = \ket{ \{  h \bx_i \},  u}_S
\end{equation}
for any $h \in C_u$. 
Intuitively, the elements in the centralizer of a permutation $u$ are all the one-to-one mappings of $N$ numbers which preserve the permutation structure of $u$. Recall that every permutation can be expanded uniquely as a product of disjoint cycles. In this representation, the centralizer is generated by $S_N$ elements which cyclically permute each disjoint cycle, as well as by permutations which swap factors with equal length. 
For example, the centralizer of the permutation $u = (1\,2)(3\,4)$ is generated by $(1\,2)$ (cyclic permutation of the first cycle), $(3\,4)$ (cyclic permutation of the second cycle), and $(1\,3)(2\,4)$ (swaps the cycles). In general, the centralizer of a permutation with $N_i$ disjoint cycles of length $i$ is isomorphic to 
\begin{equation}
\label{eq: centralizer}
\prod_i \left( S_{N_i}\ltimes \mathbb{Z}_{i}^{N_i} \right)~.
\end{equation}
Taking the redundancies \eqref{Discrete:VertexRedundancy}
into account, the inner product between symmetric states straightforwardly follows from \eqref{Discrete:InnerProductGlobalSpace}:
\begin{equation}
\label{Discrete:InnerProductSymSpace}
\sideset{_{S}}{_{S}}{\mathop{\braket{ \{ \by_i \} , v | \{ \bx_i \}, u }}} = \delta_{u, v} \sum_{h \in C_u} \left[ \prod_{i=1}^L \delta^{(N)}\left( \by_i - h\bx_i \right) \right] ~ .
\end{equation}
Note that one could have expected a Kronecker delta for the conjugacy classes of $u$ and $v$ appearing in \eqref{Discrete:InnerProductSymSpace}, rather than the group elements $u$ and $v$ themselves. However, since we picked a specific representative of each conjugacy class to label the symmetric states, one finds instead  that the inner product vanishes except when $u = v$. 

Let us take a moment to make contact with the conventional language used in symmetric product orbifold theories. 
First of all, as already mentioned, orbifold theories are typically defined in the continuum, so we can view the variables $\bX_i$ as a discretization of the continuum fields on the circle, $\bX(\phi)$. While our starting point for describing the states in the unsymmetrized theory was \eqref{Discrete:BasisStatesEigenvalues}, in the continuum one often starts with the states \eqref{Discrete:StatesInGlobalSpace} which make up $\cH_{gl}$ rather than $\cH$. The orbifold theory $\cH_S$ is then obtained by identifying global states which differ by a global $S_N$ transformation, as was done in \eqref{Discrete:GlobalSymmetryzation}. 
The local $S_N$ gauge symmetry is often implicitly assumed to be present, but rarely discussed in detail since the continuity conditions \eqref{Discrete:ContinuityConditionGeneral} may feel slightly awkward in the continuum limit. In contrast, the continuity conditions \eqref{Discrete:ContinuityConditionGlobal} imply standard continuity of the fields $\bX(\phi)$ along the circle, except at $\phi = 2\pi$ where different strands are glued into long strings, i.e., sets of strands which form a closed loop in target space according to the $S_N$ element $u$. Ultimately, the different terms in the symmetrized states \eqref{Discrete:GlobalSymmetryzation} or \eqref{Discrete:GlobalSymmetryzationCosets} correspond to distinct labelings of the same configuration of closed strings in the target space. Our step-by-step presentation, which starts with the introduction of the fully unsymmetrized states \eqref{Discrete:BasisStatesEigenvalues}, makes the origin of the joint transformation of the vertices and link variables under a permutation manifest. This, in turn, allowed us to recover the one-to-one correspondence between twisted sectors and conjugacy classes of $S_N$ from the lattice perspective, while it is usually argued for based on modular invariance in continuous orbifold CFTs \cite{Dixon:1985jw,Ginsparg:1988ui}.

Having derived a generating set of states for the symmetric Hilbert space $\cH_S$, let us now turn to more general gauge-invariant states and wavefunctions. The redundancy \eqref{Discrete:VertexRedundancy} has an important consequence for wavefunctions of gauge-invariant states of the orbifold theory. For simplicity, we shall focus on states belonging to a single twisted sector specified by $u \in \cC(S_N)$.\footnote{This discussion can be extended to linear combinations of states with support in different twisted sectors.} Expanding a symmetric state in terms of the $\ket{ \{ \bx_i \}, u }_S$:
\begin{align}
\label{Discrete:WavefunctionSymmetry}
\nonumber \ket{\Psi_u} & =  \frac{1}{\sqrt{|C_u|}} \int \dx \Psi_u (\{\bx_i\}) \ket{ \{ \bx_i \}, u }_S \\ 
\nonumber & = \frac{1}{\sqrt{|C_u|}} \int \dx \Psi_u (\{\bx_i\}) \frac{1}{|C_u|}\sum_{h \in C_u} \ket{ \{  h \bx_i \},  u  }_S \\
 & = \frac{1}{\sqrt{|C_u|}} \int \left( \prod_{i=1}^L \prod_{a=1}^N \rd x^a_i \right) \left( \frac{1}{|C_u|}\ \sum_{h \in C_u} \Psi_u (\{h^{-1}  \bx_i\}) \right) \ket{ \{ \bx_i \}, u }_S ~ ,
\end{align}
where the coefficient in front is fixed by requiring the wavefunction to be normalized. We have used the orthogonality of permutation matrices in the last equality, and integrals are over all range of $x_i^a$ variables. The final expression implies that only the totally symmetric part of a wavefunction, with respect to the centralizer $C_u$, is relevant to define a state. The wavefunctions in the symmetric Hilbert space can therefore be chosen to satisfy
\begin{equation}
\label{Discrete:WavefunctionAllowed}
\Psi_u (\{\bx_i\}) = \Psi_u (\{h  \bx_i\})
\end{equation}
for every element $h$ in the centralizer of $u$. As our notation emphasizes, the wavefunction knows about the gluing of the strands into long strings as indicated by $u$, which is the element chosen in the gauge-fixing procedure described around \eqref{Wavefunction:BasisGaugeInvariantSubspace} to label the conjugacy class (i.e., the twisted sector). This is also manifest from the fact that $\Psi_u$ is symmetric under the action of $C_u$. Had we chosen a different representative for the conjugacy class, this symmetrization would be different. Take as a working example a seed theory on a single lattice point and consider 3 copies of that theory, where the twisting is specified by $u = (1\,2) \in S_3$. The centralizer is $C_{(1\,2)} = \{ e, (1\,2) \}$, with $e$ the identity permutation. According to \eqref{Discrete:WavefunctionAllowed}, the wavefunction can be chosen to satisfy $\Psi_{(1\,2)}(x^1, x^2, x^3) = \Psi_{(1\,2)}(x^2, x^1, x^3)$. Had we chosen $(1\,3)$ to label the same twisted sector, we would have $\Psi_{(1\,3)}(x^1, x^2, x^3) = \Psi_{(1\,3)}(x^3, x^2, x^1)$.

This constraint on the gauge-invariant wavefunctions is analogous to the fact that, for identical bosonic particles in quantum mechanics, only completely symmetric wavefunctions are allowed. In one-dimensional quantum mechanics, an $N$-particle state can be expanded in unit-normalized basis states $\ket{x_1, x_2,\dots,x_N}$, with $x_1, x_2, \dots, x_N \in \mathbb{R}$. When dealing with identical bosons, one projects the Hilbert space into the subspace carrying the trivial representation of the $S_N$ symmetry that permutes the $N$ particles. This can be achieved with the projector $(N!)^{-1} \sum_{g \in S_N} \hat{g}$, which defines the symmetric Hilbert (sub)space spanned by
\begin{equation}
\label{IDparticles:SymStates}
\ket{x_1, x_2,\dots,x_N}_S = \frac{1}{\sqrt{N!}} \sum_{g \in S_N} \ket{x_{g(1)}, x_{g(2)},\dots, x_{g(N)}} ~ .
\end{equation}
Gauge-invariant states can be expanded in symmetrized states,
\begin{equation}
\label{IDparticles:GaugeInvPsiState}
\ket{\psi_S} = \frac{1}{\sqrt{N!}} \int \left( \prod_{i=1}^N \rd x_i \right) \psi_S(x_1, x_2,\dots,x_N) \ket{x_1, x_2,\dots,x_N}_S ~ ,
\end{equation}
where only fully symmetric wavefunctions are allowed:
\begin{equation}
\label{IDparticles:SymWavefunction}
\psi_S(x_1, x_2,\dots,x_N) = \psi_S(x_{g(1)}, x_{g(2)},\dots, x_{g(N)}) ~ ,
\end{equation}
for any $g\in S_N$. For future reference, we note that due to the permutation symmetry of the wavefunction of gauge-invariant states, the overlaps with symmetrized and unsymmetrized states are equal (up to normalization):
\begin{equation}
\label{IDparticles:OverlapsWavefunction}
(N!)^{-1/2} \sideset{_S}{}{\mathop{\braket{x_1, x_2,\dots,x_N | \psi_S}}} =  \psi_S(x_1, x_2,\dots,x_N) = \braket{x_1, x_2,\dots,x_N | \psi_S} ~ ,
\end{equation}
where we used that the inner product for symmetric states satisfies
\begin{equation}
\label{IDparticles:InnerProdSymmetricStates}
\sideset{_S}{_S}{\mathop{\braket{y_1,y_2, \dots, y_N | x_1,x_2, \dots, x_N}}} = \sum_{g \in S_N} \prod_{i=1}^N \delta(y_i - x_{g(i)})   ~ .
\end{equation}
We point out that many of the formulas derived above for the orbifold theory can be reinterpreted for the description of $N$ identical bosonic particles provided we restrict to the twisted sector corresponding to the identity element, $u=e$. One can make this connection precise by restricting each strand to a single site which is taken to describe the position of one of the $N$ identical particles, such that the seed theory target space is $M=\mathbb{R}$. This, in combination with $u=e$, essentially removes the continuity requirements \eqref{Discrete:ContinuityConditionGlobal}. It is then instructive to compare the constraints on the wavefunctions in both contexts. While the wavefunction of a state belonging to a non-trivial twisted sector in the orbifold theory needs only to be symmetrized over its centralizer, as per \eqref{Discrete:WavefunctionAllowed}, the wavefunction of $N$ identical particles needs to be symmetrized over the complete $S_N$ group, since the centralizer of the identity element in $S_N$ is the group itself.

%%%%%%%%%%%%%%%%%%%%%%%%%%%%%%%%%%%%%%%%
\section{Entwinement as entanglement in symmetric product orbifold theories}
\label{sec:EntwinementAsEntanglement}
%%%%%%%%%%%%%%%%%%%%%%%%%%%%%%%%%%%%%%%%

While the original proposal for entwinement as a CFT quantity dual to the length of the non-minimal, extremal geodesics in AdS$_3/\mathbb{Z}_n$ is very natural from the perspective of the quotiented geometry, the lack of a gauge-invariant formulation represents an apparent shortcoming. As mentioned in the introduction, several works have by now addressed this issue and approached the problem from a variety of angles \cite{Balasubramanian:2016xho,Lin:2016fqk,Balasubramanian:2018ajb,Erdmenger:2019lzr}, some of which resulted in unexpected and conflicting interpretations. In this section, we revisit the algebraic approach of \cite{Balasubramanian:2018ajb}, which aimed at finding a gauge-invariant reduced density matrix in the D1/D5 symmetric product orbifold theory whose von Neumann entropy would reproduce entwinement in the CFT states dual to a conical defect and massless BTZ black holes. In that work, the authors were led to conclude that the reduced density matrices they constructed (for a number of connected strands larger than one) are dominated by configurations of disconnected strands resulting from imposing an $S_N$ symmetrization, and that the corresponding von Neumann entropy is distinct from entwinement. It was suggested that if a formulation were found where the connectedness of a subset of the stands was naturally integrated in the symmetrization, the von Neumann entropy of the resulting density matrix would produce entwinement. A clue for the resolution was implicitly suggested in section~6 of \cite{Erdmenger:2019lzr}, where a tension between the approach in \cite{Balasubramanian:2018ajb} and the notation of \cite{Erdmenger:2019lzr} was raised. Relying on the formalism laid out in section~\ref{sec:Orbifolds as discrete gauged theories}, we resolve this tension and improve on the results from \cite{Balasubramanian:2018ajb} in the present section. Concretely, we first derive gauge-invariant reduced density matrices for general bi-partitions of the variables $\bX_i$ in the discretely gauged lattice description of symmetric product orbifold theories in section~\ref{subsec:ReducedDensityMatrixEntwinement}. Subsequently, we show in section~\ref{subsec:DefinitionsOfEntwinement} that the von Neumann entropy of that density matrix reproduces entwinement, as defined in \cite{Balasubramanian:2014sra}, for suitable choices of the subset $A$. We also discuss similarities and differences with previous definitions of entwinement.

Our approach will be based on the following observation. In setups where the Hilbert space factorizes, reduced density matrices are uniquely defined by the choice of subregion \cite{nielsen2010quantum}. This follows from the requirement that the reduced density matrix should contain the necessary information to reproduce expectation values for the subalgebra of operators acting on the subregion of interest. It was argued in \cite{Balasubramanian:2018ajb} that the set of gauge-invariant operators acting on subregions relevant to entwinement forms a linear subspace of operators rather than a subalgebra.\footnote{Interestingly, it was suggested in \cite{Lin:2016fqk} that there does exist a natural subalgebra associated to entwinement. We will comment on this further in section~\ref{subsec:DefinitionsOfEntwinement}.} Therefore, our goal shall be to construct a density matrix supported on $A$, which encodes the expectation values of all gauge-invariant operators acting trivially on the complement of $A$.

%%%%%%%%%%%%%%%%%%%%%%%%%%%%%%%%%%%%%%%%
\subsection{The reduced density matrix for gauge-invariant states}
\label{subsec:ReducedDensityMatrixEntwinement}
%%%%%%%%%%%%%%%%%%%%%%%%%%%%%%%%%%%%%%%%

In situations where the Hilbert space factorizes as $\cH = \cH_A \otimes \cH_{\bar{A}}$ for any spatial subregion $A$, the partial trace operation on a state described by $\rho$ has a well-understood interpretation. It produces the \textit{unique} operator on $\cH_A$, the reduced density matrix $\rho_A$, which contains all the information about $\rho$ that is localized in $A$ (see, e.g., Box 2.6 in \cite{nielsen2010quantum}). More specifically, it correctly reproduces expectation values for every observable acting only on subsystem $A$, 
\begin{equation}
\label{eq: expvalue rhoA}
    {\rm Tr}_{\cH}[\rho \, \left( \mathcal{O}_A \otimes \mathbf{1}_{\bar{A}}\right) ] = {\rm Tr}_{\cH_A}[\rho_A \mathcal{O}_A] ~ .
\end{equation}
The set of all operators on $\cH_A$ naturally forms an algebra, to which the reduced density matrix $\rho_A$ belongs.

This simple picture becomes more involved in the presence of gauge symmetries, as they generically prevent the factorization of the Hilbert space whenever the gauge transformations relate degrees of freedom in $A$ and $\bar{A}$. In view of the similarities with our discrete orbifold model, it will prove useful to first review the role of density matrices in a system of $N$ identical bosons as a warm-up, as was done in \cite{Balasubramanian:2018ajb}. We want to elucidate whether there exists some notion of a density matrix that computes expectation values for operators that act only on a subset of $k$ particles. In terms of the states \eqref{IDparticles:SymStates}, a generic symmetric $k$-particle operator is of the form
\begin{align}
\label{Operators:kParticleSymmetricVersion}
\nonumber \cO^{(k)} = \frac{1}{N!}
\int \cO_k(x_1, \dots, x_k; y_1, \dots, y_k) & \ket{x_1, \dots, x_k, z_{k+1}, \dots, z_{N}}_S \\
& \times \sideset{_S}{}{\mathop{\bra{y_1, \dots, y_k, z_{k+1}, \dots, z_{N}}}} ~ ,
\end{align}
where all the variables appearing in this expression are integrated over. The non-factorization of the symmetric Hilbert space prevents these operators to act on a tensor factor of the Hilbert space. Moreover, as already noted in \cite{Balasubramanian:2018ajb}, the operators \eqref{Operators:kParticleSymmetricVersion} do not form an algebra: the set is closed under addition, but not under multiplication. This can easily be seen by expanding the symmetric basis states as in \eqref{IDparticles:SymStates}. This produces a linear combination of terms, each of them acting on a different subset of $k$ out of $N$ particles. Multiplying two of these operators creates cross-terms that act on more than $k$ particles, which cannot be packed into an operator of the form \eqref{Operators:kParticleSymmetricVersion}. In conclusion,  the gauge symmetry represents an obstacle to having a mathematical structure associated to a subset of degrees of freedom stronger than a linear subspace of operators closed under Hermitian conjugation.

Nevertheless, expectation values of the operators $\cO^{(k)}$ in symmetric states \eqref{IDparticles:GaugeInvPsiState} can be written in terms of
a matrix that is interpreted as a reduced density matrix for $k$ identical particles:
\begin{equation}
\label{Operators:kParticleExpectationValue}
\braket{\psi_S | \cO^{(k)} | \psi_S} = \int \cO_k(x_1, \dots, x_k; y_1, \dots, y_k) \rho_S(y_1, \dots, y_k; x_1, \dots, x_k) ~ ,
\end{equation}
with
\begin{align}
\label{Wavefunction:QMDensityMatrix}
\nonumber \rho_S(x_1,\dots,x_{k};x'_1,\dots,x'_{k}) = \int \left( \prod_{i=k+1}^N\rd y_{i} \right) 
& \psi_S(x_1,\dots, x_k,y_{k+1},\dots,y_N) \\
& \times\psi_S^*(x'_1,\dots, x'_k,y_{k+1}, \dots, y_N) ~ .
\end{align}
The matrix $\rho_S$ is hence referred to as $k$-particle reduced density matrix. Although, in contrast to $\rho_A$ in \eqref{eq: expvalue rhoA}, $\rho_S$ does not act on some tensor factor of the original Hilbert space, it can be understood as encoding the state of the system for an observer that has only access to $k$ of the particles. 
As \eqref{Operators:kParticleExpectationValue} shows, knowledge of $\rho_S$ is enough to determine expectation values for all the $k$-particle operators of the form \eqref{Operators:kParticleSymmetricVersion}. Therefore, the von Neumann entropy of $\rho_S$ provides a measure for the entanglement between a subset of $k$ particles and its complement.

It is also instructive to note that \eqref{Wavefunction:QMDensityMatrix} can be obtained from a partial trace-like operation. Recall that in a factorizing Hilbert space $\cH = \cH_A \otimes \cH_{\bar{A}}$ with basis $\{\ket{\phi^{\bar{A}}_i,\phi^{A}_j}\}=\{\ket{\phi^{\bar{A}}_i}\otimes \ket{\phi^{A}_j}\}$, the matrix elements of the reduced density matrix on $\cH_A$, obtained from conventional partial tracing, are simply
\begin{equation}
\label{eq: partial trace}
\bra{\phi^A_k} \rho_{A} \ket{\phi^A_l} = \bra{\phi^A_k}{\rm Tr}_{\cH_{\bar{A}}} \left[ \rho \right]\ket{\phi^A_l} = \sum_j \bra{\phi^A_k,\phi^{\bar{A}}_j} \rho \ket{\phi^{\bar{A}}_j,\phi^A_l} ~ .
\end{equation}
The identification and sum over labels for the degrees of freedom in $\bar{A}$ in \eqref{eq: partial trace} is reminiscent of \eqref{Wavefunction:QMDensityMatrix} and suggests that the matrix elements of $\rho_S$ can be obtained by an operation mimicking \eqref{eq: partial trace}. Indeed, one can write
\begin{align}
\label{Wavefunction:QMDensityMatrix1}
\nonumber \rho_S(x_1,\dots,x_{k};x'_1,\dots,x'_{k}) = \frac{1}{N!}  \int \rd y_{k+1}\dots & \rd y_{N} \sideset{_S}{}{\mathop{\braket{x_1,\dots, x_k,y_{k+1},\dots,y_N| \psi_S}}} \\
& \quad \times \sideset{}{_S}{\mathop{\braket{\psi_S|x'_1,\dots, x'_k,y_{k+1},\dots,y_N}}} ~ .
\end{align}
Note that for this approach to be sensible, the density matrix resulting from the ``partial trace'' operation should have unit trace in the reduced space. This can be guaranteed by normalizing the states appearing in \eqref{Wavefunction:QMDensityMatrix1} such that they provide a resolution of the identity on the symmetric subspace:
\begin{equation}
\label{Wavefunction:QMIdentityOperator}
\mathbf{1}_S = \frac{1}{N!} \int \left( \prod_{i=1}^N \rd x_i \right) \ket{x_1, x_2,\dots,x_N}_S \sideset{_S}{}{\mathop{\bra{x_1, x_2,\dots,x_N}}} ~ .
\end{equation}
This motivates the $1/N!$ in \eqref{Wavefunction:QMDensityMatrix1}, which accounts for the overcompleteness of the symmetric states in the symmetric Hilbert space. 

Let us now turn to the discrete symmetric product orbifold model and introduce some new notation for later convenience. In section~\ref{sec:Orbifolds as discrete gauged theories}, we have been denoting the set of all vertex variables as $\{\bx_i\}$, where $i$ labels the different vertices on the discrete circle, each of which contains $N$ variables. In the following, we shall partition this set of degrees of freedom \emph{arbitrarily} in two groups. In particular, the bi-partition need not be  spatially organized on the lattice. We let $A$ denote an arbitrary subset of the degrees of freedom on the vertices, and $\barA$ the complementary set. Then, $\bx_A$ ($\bx_{\bar{A}}$) refers to all vertex variables in the set $A$ ($\barA$), where we drop the braces used previously to ease notation. 
To denote the full set of variables formed by combining $\bx_A$ and $\bx_{\barA}$, we use round brackets, $(\bx_A, \bx_{\barA}) = \bx$. 
Notice that it only makes sense to act with $S_N$ transformations on these complete sets of variables, an operation that we will write as $g(\bx_A, \bx_{\barA})$ or $g \bx$, where $g$ is implicitly understood to act globally (i.e., the same $g$ transformation at each lattice point). Expressions like $g \bx_A$ do not a priori make sense because $A$ is not a spatial partition in general, and therefore there may be lattice points for which only some of the $N$ degrees of freedom belong to $A$. Nevertheless, it is permitted to first act with $g \in S_N$ on a full set of vertex variables, and then apply the bi-partition defined by $A$. This operation is denoted as $(g \bx)_A$, so that $g \bx = ( (g \bx)_A, (g \bx)_{\barA})$. Finally, the integration measure will be written as a subscript.

In analogy with the operators \eqref{Operators:kParticleSymmetricVersion} in the context of identical particles, we start by considering general gauge-invariant operators that only act on a subset $A$ of the vertex degrees of freedom,
\begin{equation}
\label{Operators:SymmetricOperatorConnectedness}
\cO^{(A)} = \frac{1}{|C_u|} \int_{\bx, \by} \cO_{A}(\by_A; \bx_A) \delta(\by_{\barA} - \bx_{\barA}) \ket{\by, u}_{S} \sideset{_{S}}{}{\mathop{\bra{\bx, u}}} ~ .
\end{equation}
Notice that the subset $A$ is defined with respect to the long string configuration specified by the link variable $u$, up to the $C_u$ redundancy built in the symmetric states. In other words, we could equivalently write a symmetrized version of the operator \eqref{Operators:SymmetricOperatorConnectedness},
\begin{equation}
\label{Operators:SymmetricOperatorCuSymmetrized}
\cO^{(A)} = \frac{1}{|C_u|^2} \int_{\bx, \by} \left( \sum_{h \in C_u} \cO_{A}((h\by)_A; (h\bx)_A) \delta((h\by)_{\barA} - (h\bx)_{\barA}) \right) \ket{\by, u}_{S} \sideset{_{S}}{}{\mathop{\bra{\bx, u}}} ~ .
\end{equation}
Let us clarify the action of that operator with a simple example. Consider a system of three strands with an $S_3$ gauge symmetry, in the twisted sector labeled by $u = (1 \, 2)$. The centralizer is $C_{u} = \{e, (1\,2)\}$. If $A$ delineates the first strand and half of the second one, symmetrization by $(1\,2) \in C_u$ produces a term where the variables within $\cO_A$ are those of the second strand and half of the first one. Therefore, the statement that $A$ comprises the \emph{first} strand and half of the \emph{second} is not gauge-invariant. It is, however, perfectly legitimate to say that it selects one and a half strands within the long string of length $2$, since the $C_u$-symmetrization does not produce disconnected pieces out of originally connected ones. If we were to expand the symmetric basis states into global ones, as in \eqref{Discrete:GlobalSymmetryzationCosets}, different link variables within the conjugacy class $[u]$ would appear, but the vertex variables would also be reshuffled in a way that all terms maintain the original long string connection.

Next, we can compute expectation values of operators of the form \eqref{Operators:SymmetricOperatorConnectedness} in symmetric states \eqref{Discrete:WavefunctionSymmetry}, obtaining
\begin{equation}
\label{Operators:ExpectationValueConnectedness}
\braket{\Psi_u | \cO^{(A)} | \Psi_u} = \int_{\bx_A, \by_A} \hspace{-1em} \cO_A(\by_A; \bx_A) \rho_{S}(\bx_A; \by_A) ~ ,
\end{equation}
with
\begin{equation}
\label{Tracing:DensityMatricesSym}
\mathbf{\rho}_{S} (\bx_A, \bx'_A) =  \int_{\by_{\barA}} \Psi_u(\bx_A, \by_{\barA}) \Psi_u^{\star}(\bx'_A, \by_{\barA}) ~ ,
\end{equation}
where we used \eqref{Discrete:InnerProductSymSpace}. 
The matrix $\rho_S$ is therefore naturally interpreted as a reduced density matrix for the operators \eqref{Operators:SymmetricOperatorConnectedness} associated with the subset $A$.
Given the following resolution of the identity on the symmetric space,
\begin{equation}
\label{Tracing:ResolutionsIdentitySym}
\mathbf{1}_S  = \frac{1}{|C_u|} \int_{\bx} \ket{\bx, u}_S \sideset{_{S}}{}{\mathop{\bra{\bx, u}}} ~ ,
\end{equation}
the density matrix $\rho_S$ can also be heuristically understood as originating from a partial trace operation on $\rho = \ket{\Psi_u}\bra{\Psi_u}$, where one sums over a subset of the labels in the generating set of states,
\begin{equation}
\label{Tracing:DensityMatricesSym1}
\mathbf{\rho}_{S} (\bx_A, \bx'_A) =  \frac{1}{|C_u|} \int_{\by_{\barA}} \sideset{_{S}}{}{\mathop{\braket{(\bx_A, \by_{\barA}), u | \Psi_u}}} \braket{\Psi_u | (\bx'_A, \by_{\barA}), u}_{S} ~ .
\end{equation}
Here again, the operators \eqref{Operators:SymmetricOperatorConnectedness} do not generally close into an algebra, though they form a linear space closed under Hermitian conjugation. As we shall discuss below, the linear subspace becomes an algebra for certain choices of the subset $A$, but this is not the generic situation. 

It is interesting to compare \eqref{Tracing:DensityMatricesSym} with the analogous density matrix for identical particles \eqref{Wavefunction:QMDensityMatrix}. Although they look very similar, there is a crucial difference in their interpretation. While it is impossible to specify which of the identical particles were traced out in the latter case,  \eqref{Tracing:DensityMatricesSym} captures some information about the location of the subset $A$ inside the long string configuration specified by $u$.  
The key difference between the expressions \eqref{Wavefunction:QMDensityMatrix} and \eqref{Tracing:DensityMatricesSym} resides in the symmetrizations \eqref{IDparticles:SymWavefunction} and \eqref{Discrete:WavefunctionAllowed} imposed on the two wavefunctions. 
For identical particles, there is an $S_N$ symmetry built in the wavefunction which washes out any information in \eqref{Wavefunction:QMDensityMatrix} about which of the $N$ particles were traced out. In contrast, for a non-trivial twisted sector labeled by $u$, the wavefunction is only symmetrized over the centralizer $C_u$, as per \eqref{Discrete:WavefunctionAllowed}. Therefore, in \eqref{Tracing:DensityMatricesSym} the subset $A$ has a precise interpretation within the long string setup defined by $u$. Indeed, due to the connectedness imposed by the link $u$, the only symmetrization of $A$ that needs to be imposed within the $N$ strands is either cyclic permutations of the strands within a single long string, or the exchange of same length long strings, as understood from the general form of the centralizer \eqref{eq: centralizer}. As a consequence, in the orbifold picture it is meaningful to, e.g., trace out a connected subset $A$ within \textit{a} long string of a specified length, and this is what \eqref{Tracing:DensityMatricesSym} captures.
If more than one long string of that size exists, it is however not possible to specify in which of these long strings $A$ is to be considered. 
These observations emphasize an important point that was not appreciated enough in some of the past works: in contrast to identical particles, the $N$ strands in the orbifold theory are not really identical whenever $u$ is nontrivial, and this distinction becomes relevant when defining notions which can potentially wash that information away.

In conclusion, the von Neumann entropy 
\begin{equation}
    \label{eq:vonNeumannRhoS}
    S_{\rm vN}(\rho_S) = - {\rm Tr}[\mathbf{\rho}_{S} \log \mathbf{\rho}_{S} ]
\end{equation}
is interpreted as a measure for the entanglement between degrees of freedom whose inclusion in long strings of a certain length is specified by $A$ and $u$, and their complement. 
We stress that the matrix \eqref{Tracing:DensityMatricesSym} should not be regarded as defining an operator acting on the symmetric subspace, neither can it be interpreted as acting on a tensor factor of the original Hilbert space since there is no factorization. As such, the von Neumann entropy \eqref{eq:vonNeumannRhoS} is computed for the matrix \eqref{Tracing:DensityMatricesSym} and not for an operator on the symmetric subspace constructed out of these matrix elements. If convenient, one can heuristically picture the matrix \eqref{Tracing:DensityMatricesSym} as defining coefficients for an operator acting on an auxiliary space spanned by the vectors $\ket{\bx_A}$, with no reference whatsoever to the connectedness of the different variables, and where each variable in $A$ is understood to take values in the target space $M$ of the original theory. This is similar to how the coefficients $\cO_{A}(\by_A; \bx_A)$ of the operators \eqref{Operators:SymmetricOperatorConnectedness} are interpreted.

%%%%%%%%%%%%
\subsubsection*{Linear subspaces versus subalgebras}
%%%%%%%%%%%%

Define $\cM_S^A$ to be the set of all operators of the form \eqref{Operators:SymmetricOperatorConnectedness}. Let us elaborate a bit more on the question of whether or not $\cM_S^A$ is a subalgebra of gauge-invariant operators. The product of $\cO^{(A)},\cQ^{(A)} \in \mathcal{M}^A_S$ for a given choice of $A$ can be written as
\begin{align}
\label{eq: product of gauge invariant operators}
\nonumber \cO^{(A)} \cQ^{(A)} = \frac{1}{|C_u|^2} \int_{\bx, \by, \bz} \sum_{h \in C_u} & \cO_{A}(\by_A; (h\bz)_A) \cQ_{A}(\bz_A; \bx_A) \\
& \times \delta(\by_{\barA} - (h \bz)_{\barA}) \delta(\bz_{\barA} - \bx_{\barA}) \ket{\by, u}_{S} \sideset{_{S}}{}{\mathop{\bra{\bx, u}}} ~ .
\end{align}
In general, such a product cannot be manipulated into the form \eqref{Operators:SymmetricOperatorConnectedness}, meaning that $\mathcal{M}^A_S$ is at most a linear subspace closed under Hermitian conjugation. 

However, there are well-known situations where $\mathcal{M}^A_S$ should close into a subalgebra, namely for spatial bi-partitions where $A$ includes all $N$ vertex variables of a selection of lattice points. The key observation for spatial partitions is that any element $h \in C_u$ reshuffles elements in $A$ and $\barA$, respectively, but there is no mixing between the two subsets. In the absence of mixing, one can constrain the matrix elements of $\cO^{(A)}$ in \eqref{Operators:SymmetricOperatorCuSymmetrized}. If $h$ does not mix $A$ and $\barA$, we can show that $\delta\left( (h\bx)_{\barA} - (h\by)_{\barA} \right) = \delta\left( h\bx_{\barA} - h\by_{\barA} \right) = \delta\left( \bx_{\barA} - \by_{\barA} \right)$, and then the symmetrization only affects the reduced matrix element $\cO_A$:
\begin{equation}
\label{Operators:SymmetricOperatorCuSymmetrized2}
\cO^{(A)} = \frac{1}{|C_u|^2} \int_{\bx, \by} \left( \sum_{h \in C_u} \cO_{A}(h\by_A; h\bx_A) \right) \delta(\by_{\barA} - \bx_{\barA}) \ket{\by, u}_{S} \sideset{_{S}}{}{\mathop{\bra{\bx, u}}} ~ .
\end{equation}
As a consequence, without loss of generality, we can take the matrix elements on $A$ as $C_u$ invariant, $\cO_A(h\bx_A; h\by_A) = \cO_A(\bx_A; \by_A)$. Then, the product of two operators can be rewritten as 
\begin{align}
\cO^{(A)}\cQ^{(A)} & = \frac{1}{|C_u|^2} \sum_{h \in C_u} \int_{\bx, \by, \bz_A}\hspace{-1em}  \cO_{A}(\by_A; h\bz_A) \cQ_{A}(\bz_A; \bx_A) \delta(\by_{\barA} - h \bx_{\barA})  \ket{\by, u}_{S} \sideset{_{S}}{}{\mathop{\bra{\bx, u}}} \nonumber \\
&= \frac{1}{|C_u|^2} \sum_{h \in C_u} \int_{\bx, \by, \bz_A}\hspace{-1em}  \cO_{A}(h\by_A; h\bz_A) \cQ_{A}(\bz_A; \bx_A) \delta(h\by_{\barA} - h \bx_{\barA})  \ket{\by, u}_{S} \sideset{_{S}}{}{\mathop{\bra{\bx, u}}} \nonumber \\
&= \frac{1}{|C_u|^2} \sum_{h \in C_u} \int_{\bx, \by, \bz_A}\hspace{-1em}  \cO_{A}(\by_A; \bz_A) \cQ_{A}(\bz_A; \bx_A) \delta(\by_{\barA} - \bx_{\barA})  \ket{\by, u}_{S} \sideset{_{S}}{}{\mathop{\bra{\bx, u}}} \nonumber \\
&= \frac{1}{|C_u|} \int_{\bx, \by, \bz_A}\hspace{-1em}  \cO_{A}(\by_A; \bz_A) \cQ_{A}(\bz_A; \bx_A) \delta(\by_{\barA} - \bx_{\barA})  \ket{\by, u}_{S} \sideset{_{S}}{}{\mathop{\bra{\bx, u}}} ~ .
\label{eq:subalgebra}
\end{align}
This is of the form \eqref{Operators:SymmetricOperatorConnectedness} and therefore, whenever the centralizer of $u$ does not mix $A$ and $\bar{A}$, $\mathcal{M}^A_S$ closes into an algebra. This condition is in particular satisfied for spatial separations in any twisted sector, and this in turn allows us to recover the well-known fact that the reduced density matrix for spatially separated degrees of freedom computes expectation values for a subalgebra of operators.

%%%%%%%%%%%%%%%%%%%%%%%%%%%%%%%%%%%%%%%%
\subsection{Definitions of entwinement in the D1/D5 orbifold CFT}
\label{subsec:DefinitionsOfEntwinement}
%%%%%%%%%%%%%%%%%%%%%%%%%%%%%%%%%%%%%%%%

In the previous section, we constructed a reduced density matrix associated to a linear subspace of operators that act nontrivially on a (not necessarily spatially organized) subset of the gauged degrees of freedom $A$, in general symmetric product orbifold theories. Let us now concentrate on the D1/D5 orbifold CFT which emerges in the ${\rm AdS}_3$/${\rm CFT_2}$ correspondence discussed in the introduction, and show that the von Neumann entropy of $\rho_S$ reproduces entwinement. Several notions and interpretations of entwinement have been offered since the original proposal, some of which are at odds with or diverge from our developments. In this section, we provide an overview of alternative methods with a focus on similarities and differences with the formalism described above. 

Let us first recall that within the D1/D5 orbifold CFT, the state dual to the conical defect obtained as a $\mathbb{Z}_n$ quotient of AdS$_3$ can be schematically written as \cite{Balasubramanian:2000rt,Martinec:2001cf,Martinec:2002xq,Balasubramanian:2005qu}
\begin{equation}
\label{eq: conical defect CFT state}
    \ket{\Psi_u} = \left[\sigma_{n}(0)\right]^{N/n} \ket{0} ~ .
\end{equation}
The action of the twist operator $\sigma_{n}^{N/n}$ effectively glues the $N$ strands into $N/n$ long strings of length $n$. Each of the long strings settles in the vacuum state of the CFT defined on a circle that is $n$ times the size of the original circle. In \eqref{eq: conical defect CFT state}, we assume that the RHS has already been symmetrized over the symmetric group. For concreteness, a representative for the corresponding twisted sector can be chosen as
\begin{equation}
\label{eq: u}
    u = (1 \, 2\dots n)(n\!+\!1 \dots 2n) \dots (N\!-\!n\!+\!1 \dots N) ~ .
\end{equation}
The target space of the orbifold theory is $(T^4)^N/S_N$.

We discussed in detail how to interpret the von Neumann entropy of the matrix $\rho_S$ for an arbitrary subset $A$ in the previous section. In the context of the state \eqref{eq: conical defect CFT state}, the original definition of entwinement relied on the intuition that the lengths of geodesics winding around the conical defect should be described by the entanglement entropy of the boundary region homologous to the said geodesic after unfolding the $\mathbb{Z}_n$ quotient \cite{Balasubramanian:2014sra}. Prior to comparing with other attempts to define entwinement, we demonstrate that the von Neumann entropy of the matrix $\rho_S$ encodes the length of winding geodesics for $A$ naturally associated with the relevant region in the unfolded boundary.

On a fixed time-slice, the quotient geometry AdS$_3/\mathbb{Z}_n$ allows for $n$ geodesics anchored to any two points separated by an angle $\alpha$ on the boundary. The different geodesics can be parameterized by an integer $\ell = 0, \dots, n-1$ and their lengths correspond to \cite{Balasubramanian:2016xho}
\begin{equation}
\label{eq: lengths winding geodesics}
\mathcal{L}_\ell (\alpha)=2 R_{\rm AdS} \log \left[ \frac{2 n r_\infty}{R_{\rm AdS}} \sin\left( \frac{\alpha+2\pi \ell}{2 n} \right) \right] ~,
\end{equation}
with $r_\infty$ an infrared regulator. 

Let us now compute the von Neumann entropy of $\rho_S$ for a suitable subset $A$ in the state \eqref{eq: conical defect CFT state} dual to AdS$_3/\mathbb{Z}_n$ and recover \eqref{eq: lengths winding geodesics} as a generalization of the RT formula \eqref{Intro:RyuTakayanagi}. The continuum limit of the discretized orbifold theory we have been discussing is a CFT of $N/n$ long strings of length $n$. The orbifold theory is defined on a circle of length $2 \pi R_{\rm AdS}$. The subset $A$ is chosen as the union of $N/n$ identical, connected pieces, one in each of the long strings, made out of a continuous portion of $l$ strands together with an additional piece of opening angle $\alpha$ in the following strand.  The wavefunction in the state \eqref{eq: conical defect CFT state} is the product of vacuum wavefunctions for each long string: $\Psi_u(\mathbf{x}) = \psi_0(\bx_{LS_1}) \dots \psi_0 (\bx_{LS_{N/n}})$, where $\psi_0$ is the vacuum wavefunction of one long string and $LS_1, \dots, LS_{N/n}$ is the partition of the degrees of freedom into each long string as dictated by \eqref{eq: u}. Thus, the full density operator $\rho$ is equivalently written as the product of $N/n$ independent density matrices for each long string. For $A$ as described above, the reduced density matrix $\rho^{(l, \alpha)}_S$, \eqref{Tracing:DensityMatricesSym}, analogously factorizes into $N/n$ independent reduced density matrices, and the resulting von Neumann entropy splits into a sum of $N/n$ equal entropies. The von Neumann entropy of $\rho^{(l, \alpha)}_S$ is hence determined by the entropy of a single long string. For $c_{ls}$ the central charge of the theory of a single long string (i.e., $c_{ls} = c \, n / N$, with $c$ the central charge of the full orbifold theory) one obtains by standard CFT methods \cite{Holzhey:1994we,Calabrese:2009qy}
\begin{equation}
\label{Comparisons:EntwinementFromFieldTheory}
S_{\rm vN} \left( \rho^{(l, \alpha)}_S \right) = \frac{N}{n} \frac{c_{ls}}{3} \log \left[ \frac{2 n R_{\rm AdS}}{\epsilon_{\rm UV}} \sin \left( \frac{\alpha + 2 \pi l}{2 n} \right) \right] ~ .
\end{equation}
We can thus check that entwinement, understood as the von Neumann entropy of $\rho^{(l, \alpha)}_S$ \eqref{eq:vonNeumannRhoS}, reproduces the length of geodesics with non-trivial winding as 
\begin{equation}
E_l(\alpha) \equiv S_{\rm vN} \left( \rho^{(l, \alpha)}_S \right) = \frac{\mathcal{L}_\ell (\alpha)}{4 G_N} ~ ,
\end{equation}
where we used the standard Brown-Henneaux relation for the central charge $N c_{ls} / n =c= 3R_{\rm AdS}/(2 G_N)$, and where the cutoffs relate as $r_{\infty} = R^2_{\rm AdS}/\epsilon_{\rm UV}$. This reproduces the conclusion of \cite{Balasubramanian:2014sra}, derived now from our definition of the symmetric reduced density matrix.

\subsubsection*{Earlier work on reduced density matrices in symmetric product orbifolds}

The problem of recasting entwinement as the von Neumann entropy of a reduced density matrix for internal, discretely gauged degrees of freedom in the orbifold description was first undertaken in \cite{Balasubramanian:2018ajb}. 
The authors found that, while for a single strand the procedure correctly reproduces entwinement, the density matrix reduced on more than a single strand receives contributions from connected and disconnected parts across the different long strings (cf.~(4.33) and (4.38), respectively, in \cite{Balasubramanian:2018ajb}), the latter type of contribution being dominant in the large $N$ limit. 
As a result, it was argued that the von Neumann entropy of the reduced density matrix for multiple strands in general differs from entwinement. In contrast, in the present work we demonstrated the existence of a gauge-invariant density matrix whose von Neumann entropy does reproduce entwinement. Since the notion of entwinement has been argued to be naturally associated with a linear subspace of operators rather than a subalgebra, our approach was rooted in the  identification of a density matrix that encodes expectation values for the linear subspace of \textit{gauge-invariant} operators (i.e., which can be expanded in terms of the gauge-invariant states \eqref{Discrete:GlobalSymmetryzation}) whose coefficients contain a delta function in the degrees of freedom in $\bar{A}$. In addition, we demonstrated that the reduced density matrix \eqref{Tracing:DensityMatricesSym} can be obtained by means of a procedure that mimics conventional partial tracing in factorizing theories. 

This second approach to obtaining \eqref{Tracing:DensityMatricesSym} is in fact very much in the spirit of \cite{Balasubramanian:2018ajb}, yet our results diverge. 
The mismatch finds its origin in our more careful implementation of the $S_N$ symmetrization: a group element $\hg$ acts simultaneously on the vertex and the link variables of a state, in such a way as to conserve the connectedness of the longs strings. As emphasized in section~\ref{sec:Orbifolds as discrete gauged theories}, our constructive approach to the symmetrized states \eqref{Discrete:GlobalSymmetryzation} makes it manifest that the operation of symmetrization merely captures distinct labelings of a same target space configuration. In other words, what is connected in the target space in one representation of a symmetrized state, stays connected in every term of the symmetrized state. 
In contrast, in the symmetrization of the wavefunction in, e.g., (4.6) and (4.13) of \cite{Balasubramanian:2018ajb},
either the link or the vertex variables were considered to transform under $\hg$, rather than transforming both in a coordinated way.\footnote{Relatedly, there is some ambiguity in the notation for the overlap of the wavefunction with a basis state in (4.5)-(4.6) of  \cite{Balasubramanian:2018ajb}, since it does not clearly label the twisted sector in which the basis state lives.}  As a consequence, initially connected portions of $A$ became disconnected under the action of the symmetrization and these configurations contributed to the reduced density matrix.

\subsubsection*{The extended Hilbert space method}
The first attempt to define entwinement as a CFT quantity dual to the length of extremal, non-minimal geodesics in AdS$_3/\mathbb{Z}_n$ \cite{Balasubramanian:2014sra} revolved around the idea of ungauging the $\mathbb{Z}_n$ symmetry of the quotiented geometry, without being specific about the $N$ internal gauged degrees of freedom. The method\footnote{Recall that the approach of \cite{Balasubramanian:2014sra} consists in embedding the initial gauge-invariant state in a larger Hilbert space where the part of the gauge-symmetry that relates different strands within a long string has been relaxed. After computing standard entanglement entropy in the presence of these additional unphysical degrees of freedom, the result is symmetrized over the $n$ copies which yields a $\mathbb{Z}_n$-invariant answer.} is inspired from a technique in lattice gauge theories known as the extended Hilbert space method \cite{Ghosh:2015iwa,Aoki:2015bsa,Soni:2015yga}. This approach is often regarded as unsatisfactory because it involves tracing over unphysical degrees of freedom which do not belong to the symmetric Hilbert space. Nonetheless, it is instructive to phrase their procedure in our formalism and demonstrate the equivalence of the resulting density matrix with our $\rho_S$ within the symmetric Hilbert space $\cH_S^u$.

The quotiented geometry is described by the twisted sector characterized by \eqref{eq: u}. The geometrical picture of ungauging the $\mathbb{Z}_n$ involves going to an ${\rm AdS}_3$ covering space which has an $n$ times longer boundary in which the long strings are unwound. In terms of the orbifold CFT, this is realized by ungauging the $\mathbb{Z}_n^{N/n}$ subgroup within the centralizer of all the elements in $[u]$ -- recall the structure of the centralizer presented in \eqref{eq: centralizer}, which for this twisted sector becomes $S_{N/n} \ltimes \mathbb{Z}_n^{N/n}$. 
Intuitively, this originates from the observation that ungauging the $\mathbb{Z}_n$-symmetry of the geometry, as in Figure~\ref{fig: geodesics conical defect}, allows us to distinguish the $m$-th strand in any of the long strings. Hence, the cyclic permutations of the long strings are no longer part of the gauge group in the covering theory. This corresponds to ungauging the $\mathbb{Z}_n^{N/n}$ subgroup. 

The symmetric states \eqref{Discrete:GlobalSymmetryzation} can be explicitly partially ungauged as follows. Recall that, for a given choice of set $K$ transversal to $C_u$, consisting of elements that relate the link variable $u$ to other link variables $v = k u k^{-1}$ in its conjugacy class, the centralizer $C_u$ is isomorphically mapped to $C_v$ as $l = k h k^{-1}$, where $h \in C_u$ and $l \in C_v$. Then, the sums over $K$ and $C_u$ in \eqref{Discrete:GlobalSymmetryzationCosets} can be traded for sums over $[u]$ and $C_v$,
\begin{equation}
\label{eq:SymmetricSplitU}
\ket{ \{ \bx_i \}, u }_{S} = \frac{1}{\sqrt{N!}} \sum_{k \in K} \sum_{h \in C_u} \ket{ \{ k h \bx_i \}, k u k^{-1}}_{gl} = \frac{1}{\sqrt{N!}} \sum_{v \in [u]} \sum_{l \in C_v} \ket{ \{ l k \bx_i \}, v }_{gl} ~ ,
\end{equation}
where $k$ is the unique element in $K$ such that $k u k^{-1} = v$. We can now define (partially) ungauged states by restricting the sums over the centralizers to the $S_{N/n}$ factor only (see \eqref{eq: centralizer}) in both expressions:
\begin{align}
\label{ExtendedSpace:PartialGauging}
\nonumber \ket{ \{ \bx_i \}, u }_{UG} & \equiv \sqrt{\frac{|C_u|}{N! (N/n)!}} \sum_{v \in [u]} \sum_{s_v \in S_{N/n}^{(v)}} \ket{ \{ s_v k \bx_i \}, v }_{gl} \\
& = \sqrt{\frac{|C_u|}{N! (N/n)!}} \sum_{k \in K} \sum_{s \in S_{N/n}} \ket{ \{ k s \bx_i \}, k u k^{-1} }_{gl} ~ .
\end{align}
In this equation, $S_{N/n}^{(v)}$ is the $S_{N/n}$ factor within $C_v$, while we refer to $S_{N/n}^{(u)}$ simply as $S_{N/n}$. Furthermore, we have used the isomorphism between the two groups provided by $s_v = k s k^{-1}$, as before. 
As pointed out in \cite{Balasubramanian:2014sra} and \cite{Erdmenger:2019lzr}, the states spanning the extended Hilbert space are expected to generate an $S_{N/n}$-symmetric product orbifold theory. This picture is clearest when interpreting the states \eqref{ExtendedSpace:PartialGauging} as living on a circle of size $2\pi n R_{\rm AdS}$, effectively unwrapping the long strings of the theory. This perspective highlights the resulting $S_{N/n}$ symmetry which simply permutes the long strings, but it is in principle not needed to verify the $S_{N/n}$ invariance of the states \eqref{ExtendedSpace:PartialGauging}. Indeed, one can explicitly construct $S_{N/n}$ elements which act trivially on \eqref{ExtendedSpace:PartialGauging}:
\begin{equation}
    \hat{g}_s = \sum_{k\in K} \hat{k}\hat{s}\hat{k}^{-1} \ket{kuk^{-1}}\bra{kuk^{-1}} ~ ,
\end{equation}
for $s\in S_{N/n}$, the elements in $S_N$ permuting the $N/n$ long strings of the twisted sector labeled by $u$.
Fully symmetrized states can be recovered from \eqref{ExtendedSpace:PartialGauging} after projecting with the $\mathbb{Z}^{N/n}_n$ factor within each $C_v$. 

Note that, in this partially ungauged picture, the subsets $A$ considered in the study of entwinement, which give rise to \eqref{Comparisons:EntwinementFromFieldTheory} (i.e., subsets $A$ that are invariant under the $S_{N/n}$ action that interchanges long strings), can be interpreted as spatial regions on the circle of size $2\pi n R_{\rm AdS}$.
Hence, for this choice of subregions $A$ the set of partially gauged operators
\begin{equation}
\cO^{(A)} = \frac{1}{(N/n)!} \int_{\bx, \by} \cO_{A}(\by_A; \bx_A) \delta(\by_{\barA} - \bx_{\barA}) \ket{\by, u}_{UG} \sideset{_{UG}}{}{\mathop{\bra{\bx, u}}}
\label{eq:ungauged operator subalgebra}
\end{equation}
closes into a subalgebra (by a reasoning similar to \eqref{eq:subalgebra}). Using $\sideset{_{UG}}{}{\mathop{\braket{\bx, u|\Psi_u} \sim \Psi_u(\bx)}}$, the expectation value of the operators \eqref{eq:ungauged operator subalgebra} in gauge-invariant states can be obtained from the reduced density matrix $\rho_S$ \eqref{Tracing:DensityMatricesSym}. In the extended Hilbert space approach, entwinement, interpreted as the von Neumann entropy of the resulting density matrix, is nothing but conventional algebraic entanglement entropy. This was of course the perspective taken in the original paper \cite{Balasubramanian:2014sra} and translated in our formalism.

Alternatively, we can reach the conclusion that the reduced density matrix on $A$ in the ungauged theory is identical to \eqref{Tracing:DensityMatricesSym} by designing a partial trace operation, analogous to \eqref{Tracing:DensityMatricesSym1}, but now based on projecting $\rho$ on the states \eqref{ExtendedSpace:PartialGauging} while integrating the degrees of freedom in $\bar{A}$:
\begin{align}
\label{Tracing:DensityMatricesUG}
\nonumber \mathbf{\rho}_{UG} (\bx_A, \bx'_A) & \equiv  \frac{1}{(N/n)!} \int_{\by_{\barA}}  \sideset{_{UG}}{}{\mathop{\braket{(\bx_A, \by_{\barA}), u | \Psi_u}}} \braket{\Psi_u | (\bx'_A, \by_{\barA}), u}_{UG} \\
& =  \int_{\by_{\barA}} \Psi_u(\bx_A, \by_{\barA}) \Psi_u^{\star}(\bx'_A, \by_{\barA})= \mathbf{\rho}_{S} (\bx_A, \bx'_A) ~ ,
\end{align}
where we used that 
\begin{equation}
\label{Tracing:InnerProductsUG}
\sideset{_{UG}}{}{\mathop{\braket{\bx, v | \Psi_u}}} = \sqrt{(N/n)!} \Psi_u (  \bx) \delta_{u,v} ~ ,
\end{equation}
and chose the normalization of the tracing procedure according to a second resolution of the identity on the symmetric subspace $\cH_S^u$:
\begin{equation}
\label{Tracing:ResolutionsIdentityUG}
\mathbf{1}_{UG}  = \frac{1}{(N/n)!}  \int_{\bx} \ket{\bx, u}_{UG} \sideset{_{{UG}}}{}{\mathop{\bra{\bx, u}}} ~ .
\end{equation}
This guarantees a unit result in case of a complete trace operation (i.e., integrating over all the degrees of freedom instead of just $\bar A$ in \eqref{Tracing:DensityMatricesUG}).
These arguments confirm that our density matrix approach to entwinement is essentially equivalent to the original, extended Hilbert space method \cite{Balasubramanian:2014sra}.

\subsubsection*{From the replica trick}

Another perspective on the notion of entwinement as a measure for the entanglement between internal gauged degrees of freedom in orbifold theories was provided in \cite{Balasubramanian:2016xho}, based on the replica trick for two-dimensional CFTs. In the standard replica trick approach for spatial entanglement entropy, the computation is geared towards obtaining Renyi entropies from the two-point correlator of replica twist fields, where the position of the two twist fields on the circle delineates the spatial region $A$ of interest. The entanglement entropy is subsequently found by analytically continuing the outcome for the $n^{\rm{th}}$ Renyi entropy to $n \rightarrow 1$. This procedure was generalized in \cite{Balasubramanian:2016xho} to subsets of internal degrees of freedom by the introduction of replica twist fields that are charged under the discrete $S_N$ gauge symmetry. This allows for the insertion of replica twist fields on single strands instead of on the spatial domain of the theory, which effectively delineates regions on the long string configurations that do \textit{not} generically correspond to spatial bi-partitions on the original orbifold CFT. The von Neumann entropy of internal degrees of freedom computed this way was found to agree perfectly with entwinement in all the known examples, and provided as such a first fully gauge-invariant definition of entwinement from the perspective of the orbifold CFT.

We point out, however, that although the final result (3.17) in \cite{Balasubramanian:2016xho} is consistent with the von Neumann entropy we find in \eqref{Comparisons:EntwinementFromFieldTheory} for the subsets $A$ relevant to holography, both approaches disagree for more general bi-partitions. Indeed, while our definition incorporates the possibility of specifying the length of the long string in which to consider a given connected portion of the strands in a gauge-invariant manner, (3.17) of \cite{Balasubramanian:2016xho} contains a sum over long strings in which to consider that portion. The origin of this discrepancy can again be found in the way the calculation in \cite{Balasubramanian:2016xho} was made gauge-invariant. Their approach consisted in computing the expectation value of two replica twist fields, inserted an angular distance $2\pi l + \alpha$ away from each other within a long string, while the position of the first twist field is $S_N$-symmetrized over all possible strands (cf.~(2.5) in \cite{Balasubramanian:2016xho}). The way it was implemented in \cite{Balasubramanian:2016xho}, this procedure had the drawback that the 
interval delineated by the twist operators always needed to be considered in every possible long string. We now present a refinement of the replica method using the following gauge-invariant insertions:
\begin{align}
\label{eq:TwistOperatorsImproved}
\sum_{g \in S_N} \hg \tilde{\Sigma}_{i}(0) \Sigma_{u^l(i)}(\alpha) \hg^{-1}\otimes \hg &\ket{u}\bra{u}\hg^{-1} ~ \nonumber \\
&=  \sum_{g \in S_N}  \tilde{\Sigma}_{g(i)}(0) \Sigma_{g u^l(i)}(\alpha) \otimes \ket{gug^{-1}}\bra{gug^{-1}} \nonumber \\
&=  \sum_{g \in S_N}  \tilde{\Sigma}_{g(i)}(0) \Sigma_{{(gug^{-1}})^lg(i)}(\alpha) \otimes \ket{gug^{-1}}\bra{gug^{-1}} ~.
\end{align}
Here, $\tilde{\Sigma}_{i}(0)$ denotes the insertion of a first twist operator at the origin of strand $i$. The region $A$ extends between the first and second twist field. The latter, $\Sigma_{u^l(i)}(\alpha)$, is inserted after going $l$ times around the circle, ending on strand $u^l(i)$, and rotating an angle $\alpha$ inside that strand. The continuity of the interval delineated by the twist operators along the long string of strand $i$, which needed to be assumed in \cite{Balasubramanian:2016xho}, is automatic in \eqref{eq:TwistOperatorsImproved} by the inclusion of the projector $\ket{u}\bra{u}$. Moreover, this also ensures that the action of a group element on an unsymmetrized representative operator in \eqref{eq:TwistOperatorsImproved} simultaneously transforms the location of the twist fields and the link variable in the projector in a coordinated way, which conserves the connectedness of the interval $A$ as well as its location within the long string configuration. This is manifest in the last form of the previous expression. In contrast to \cite{Balasubramanian:2016xho}, the symmetrization of the twist operators \eqref{eq:TwistOperatorsImproved} does not bring the interval to long strings of different lengths.
This operation merely rearranges the location of the interval according to the centralizer $C_u$. 
Once again, we find that it is sensible to consider the entanglement entropy of a segment within a long string of a certain length, but we cannot tell apart strands within a given long string, nor can we distinguish long strings of the same length.

\subsubsection*{Entwinement as the algebraic entanglement entropy of a subalgebra}

As we emphasized above, the reduced density matrix $\rho_S$ associated to entwinement encodes information about expectation values of a linear subspace of operators acting on $A$, rather than a subalgebra. This can be contrasted with standard lattice gauge theory setups where the entanglement entropy of spatial subregions has been thoroughly studied, see e.g.\ \cite{Buividovich:2008gq,Buividovich:2008yv,Donnelly:2011hn,Casini:2013rba,Radicevic:2014kqa,Donnelly:2014gva,Ghosh:2015iwa,Aoki:2015bsa,Soni:2015yga,VanAcoleyen:2015ccp}. These works include algebraic approaches to quantifying entanglement, where several prescriptions are provided for constructing gauge-invariant subalgebras corresponding to spatial subregions \cite{Casini:2013rba,Radicevic:2014kqa}. Therefore, even though the gauge-invariant Hilbert space does not admit a tensor factor decomposition for spatial bi-partitions, the existence of algebras associated to subregions allows for the derivation of reduced density matrices with associated measures for the entanglement between the subregion and its complement. 

This appealing perspective on entanglement entropy in non-factorizing Hilbert spaces inspired the conjecture of \cite{Lin:2016fqk}, which asserts the existence of a gauge-invariant density matrix associated with a gauge-invariant subalgebra of operators whose von Neumann entropy reproduces entwinement. The conjecture comes with a detailed recipe to construct the relevant gauge-invariant subalgebra, which contains the reduced density matrix as an element. 
In short, the idea is the following. Schematically, the natural generators of a subalgebra of operators associated to a non-spatial subset of degrees of freedom $A$ are of the form
\begin{equation}
\label{eq: symmetric operators on A}
\mathcal{O}_A\otimes \mathbf{1}_{\bar{A}} + \text{gauge-transformations} ~.
\end{equation}
The common issue with this suggestion is that these operator generally generate the entire algebra on $\cH_S$. The approach of \cite{Lin:2016fqk} is to consider the projection of the original density matrix $\rho$ on the linear subspace generated by the operators of the type \eqref{eq: symmetric operators on A}. The resulting operator, denoted as $\mathcal{O}_\rho$, is then taken as the sole generator of a subalgebra of gauge-invariant operators $\mathcal{A}_{\mathcal{O}_{\rho}}$, which in general is expected to close without generating the entire algebra. For the final reduced density matrix to be a proper density matrix for the ensuing subalgebra, one should define $\rho_{red}$ as the projection of $\rho$ on the subalgebra $\mathcal{A}_{\mathcal{O}_{\rho}}$. 
The non-trivial claim of \cite{Lin:2016fqk} is that the von Neumann entropy of $\rho_{red}$ is equal to entwinement defined using the extended Hilbert space method. 

The recipe we just described was verified by explicit computation in \cite{Lin:2016fqk} for a very simple model consisting of two identical spin-$1/2$ degrees of freedom with a $\mathbb{Z}_2$ gauge symmetry, with $A$ being one of the two spins. 
One can straightforwardly generalize this setup to include additional sites with an arbitrary number $N$ of spins per site on which a local $S_N$ gauge symmetry is imposed.
This model can be interpreted as a toy model for symmetric orbifold CFTs, though it is a priori not obvious how to incorporate a notion of connectedness between the spins. This comment aside, the proposed recipe was conjectured to work beyond the simple single site setup with $\mathbb{Z}_2$ symmetry.

However, when applying this
recipe to the next simplest example (i.e., three spins with an $S_3$ gauge symmetry), we found that it does not produce
a reduced density matrix with the correct von Neumann entropy.
We defer the details of this computation to Appendix~\ref{App:Algebraic entwinement in $S_3$-symmetric spin model}. We interpret this observation as support for the general picture of our developments as well as the discussions in \cite{Balasubramanian:2018ajb,Erdmenger:2019lzr}: the natural mathematical structure associated to a reduced density matrix for internal gauged degrees of freedom appears to be a linear subspace of operators rather than a subalgebra.

\subsubsection*{Entwinement from projective measurements}
More recently, another formulation of the interpretation of entwinement as a quantum information theoretic measure associated to a linear subspace of gauge-invariant operators was derived for general CFTs with a $\mathbb{Z}_n$ gauge symmetry \cite{Erdmenger:2019lzr}. In that work, the authors consider a subset of degrees of freedom in the covering theory of the $\mathbb{Z}_n$-symmetric CFT and acknowledge, in agreement with \cite{Balasubramanian:2018ajb} and the present analysis, that this subset is naturally associated with a linear subspace of gauge-invariant operators on the $\mathbb{Z}_n$-symmetric subspace.
Based on this observation, they consider the von Neumann entropy of the probability distribution resulting from a projective measurement on $\rho$, drawn from that linear subspace, and show that entwinement can be found as the minimum of these von Neumann entropies. This interpretation emphasizes the information theoretic role of the linear subspace in the definition of entwinement.

%%%%%%%%%%%%%%%%%%%%%%%%%%%%%%%%%%%%%%%%
\section{Discussion}
\label{sec:Discussion}
%%%%%%%%%%%%%%%%%%%%%%%%%%%%%%%%%%%%%%%%

The main goal of our work was to construct an explicit framework based on the familiar notion of density matrices to quantify entanglement between internal, gauged degrees of freedom in symmetric product orbifold CFTs. We have done so by formulating a lattice model, whose main advantage is to naturally introduce the notion of twisted sectors (by means of a link variable $u \in S_N$ that indicates how the strands are glued when going around the base space circle), which allowed to clearly identify the gauge transformation properties of vertex and link variables. Using this formalism, we improved on the results of \cite{Balasubramanian:2018ajb} by deriving a reduced density matrix for a general subset $A$ of the degrees of freedom in the orbifold theory, \eqref{Tracing:DensityMatricesSym}, whose von Neumann entropy agrees with the original definition of entwinement. 
We concluded with an account of prior definitions of entwinement  \cite{Balasubramanian:2016xho,Lin:2016fqk,Erdmenger:2019lzr}, emphasizing similarities and differences with our results. 

We point out that our framework is applicable to other setups than those relevant to the holographic questions that served as motivation. In particular, \eqref{Tracing:DensityMatricesSym} and \eqref{eq:vonNeumannRhoS} are defined for any twisted sector. Moreover, the subset $A$ can be defined at will for a given link $u \in S_N$. Our results essentially generalize the path to describing entanglement for identical particles to orbifold theories, by relying on the intuition that the reduced density matrix and its associated entropy should encode all the information about expectation values of operators acting solely on a restricted set of degrees of freedom. Finally, we remark that similar constructions can be developed for other gauge groups.  

Regarding future directions and connections with other works, we can broaden our scope and leave the restricted setup of symmetric product orbifold theories. In generic local quantum field theories, considering spatially separated bi-partitions for the entanglement entropy is natural, though in principle not necessary. One alternative bi-partition that has received some attention in the past is splitting the momentum space \cite{Balasubramanian:2011wt}, which measures entanglement between IR and UV degrees of freedom. More recently, there has been a series of works with a focus on partitioning the target space of a theory instead of its base space \cite{Mazenc:2019ety}. This setup appears naturally in the context of worldsheet string theory and $Dp$ brane holography \cite{Das:2020jhy,Das:2020xoa,Hampapura:2020hfg}. The fact that these works deal with non-spatial partitions in the presence of gauge symmetries suggests that some lessons learned in our present setup can potentially be fruitfully applied in these other situations as well. 
Finally, and more ambitiously, there are well-known models where a dynamical spacetime picture is found to emerge from a theory containing only matrix degrees of freedom (the BFSS model \cite{Banks:1996vh} is a paradigmatic example of this, but certainly not the only one \cite{Kazakov:2000pm}). 
In these theories with no spatial extent, it appears that the connection between emergent geometric notions in the spacetime and quantum entanglement of the fundamental degrees of freedom should necessarily be encoded in the entanglement structure of internal degrees of freedom. 
It will be interesting to see whether the ideas presented in this paper can be of use in these contexts.

\section*{Acknowledgements}

We thank Riccardo Argurio, Marius Gerbershagen, Jennifer Lin and Alfonso V. Ramallo for useful discussions. This research has been supported by FWO-Vlaanderen project G012222N and by Vrije Universiteit Brussel through the Strategic Research Program High-Energy Physics. MDC is partially supported by the Simons Foundation Award number 620869 and by STFC consolidated grant ST/T000694/1. AVL is supported by the F.R.S.-FNRS Belgium through conventions FRFC PDRT.1025.14 and IISN 4.4503.15, as well as by funds from the Solvay Family.

\appendix 
\section{Algebraic entwinement in an $S_3$-symmetric spin model}
\label{App:Algebraic entwinement in $S_3$-symmetric spin model}

We revisit the algebraic approach to entwinement proposed in \cite{Lin:2016fqk}, where the existence of a density matrix with the following properties was conjectured: 
\begin{itemize}
    \item it is an element of a subalgebra $\mathcal{A}$ of gauge-invariant operators,
    \item it computes expectation values for the subalgebra $\mathcal{A}$ (instead of a linear subspace), \item its von Neumann entropy reproduces entwinement as computed via the extended Hilbert space method.
\end{itemize}
This claim was supported by a brute-force analysis for the entwinement of a single spin $1/2$ degree of freedom in a toy model consisting of two spins with a $\mathbb{Z}_2$-symmetry. In this appendix, we treat the next simplest example and apply the recipe detailed in \cite{Lin:2016fqk} for the entwinement of one spin in a system of three spins with a $S_3$-symmetry. We find that the von Neumann entropy obtained this way does not always relate in the correct way to the von Neumann entropy of the extended Hilbert space reduced density matrix. In support of this claim, we first derive the entwinement of a single spin in this simple model using the extended Hilbert space method, where a gauge-invariant state is naturally embedded in the Hilbert space where the $S_3$ gauge symmetry has been relaxed. Next, we construct the gauge-invariant $\mathcal{A}_{\mathcal{O}_{\rho}}$ which follows from the recipe detailed in the main text (see section~\ref{subsec:DefinitionsOfEntwinement}) and show that the von Neumann entropy of the associated reduced density matrix differs from entwinement as found in the first method.

In the extended Hilbert space approach to entwinement, one considers a gauge-invariant state $\ket{\psi}$ and a set of degrees of freedom denoted by $A$. Due to the gauge constraints, the gauge-invariant Hilbert space $\mathcal{H}_S$ in which $\ket{\psi}$ lives does not admit a tensor product representation which splits degrees of freedom in $A$ and its complement $\bar{A}$. One can nevertheless consider a larger Hilbert space $\mathcal{H}_{ext}$ where the gauge constraints are lifted and trace out $\bar{A}$ in the usual way. For every symmetric state $\ket{\psi}$, there exists a natural embedding in $\mathcal{H}_{ext}$. Concretely, let us consider an $S_3$-symmetric spin $1/2$ model and embed a general gauge-invariant state in the extended, unsymmetrized Hilbert space $\mathcal{H}_{ext}$:
\begin{align}
    \ket{\psi} = & a \ket{\uparrow \uparrow \uparrow}_S + b  \ket{\downarrow \uparrow \uparrow}_S +  c  
    \ket{\uparrow \downarrow \downarrow}_S
      +
    d \ket{\downarrow \downarrow \downarrow}_S \nonumber \\
    \equiv & a \ket{\uparrow \uparrow \uparrow} + \frac{b}{\sqrt{3}} \left( \ket{\downarrow \uparrow \uparrow} + \ket{\uparrow \downarrow \uparrow} +
    \ket{\uparrow \uparrow \downarrow} \right) \nonumber \\
    & + \frac{c}{\sqrt{3}} \left( 
    \ket{\uparrow \downarrow \downarrow} + 
    \ket{\downarrow \uparrow \downarrow} + \ket{\downarrow \downarrow \uparrow} 
     \right) +
    d \ket{\downarrow \downarrow \downarrow} ~ ,
    \label{eq: gaugeinv S3 state}
\end{align}
with $a^2 + b^2 + c^2 + d^2 = 1$. We denote the associated $8 \times 8$ density matrix in the extended Hilbert space $\rho_{ext} \equiv \ket{\psi}\bra{\psi}$. When projected onto the gauge-invariant Hilbert space with basis $\{\ket{\uparrow \uparrow \uparrow}_S$, $ \ket{\downarrow \uparrow \uparrow}_S$, $ \ket{\uparrow \downarrow \downarrow}_S$, $\ket{\downarrow \downarrow \downarrow}_S \}$,
     the density matrix becomes
     \begin{equation}
         \rho^{S} = 
    \begin{pmatrix}
    a^2 & ab & ac & ad  \\
 ab & b^2 &bc & bd \\
  ac & bc & c^2 & cd \\
  ad & bd & cd & d^2
    \end{pmatrix} ~ .
    \label{eq: 4d density matrix}
     \end{equation}

The original formulation of entwinement for one of the three spins requires one to compute conventional entanglement entropy for a single spin in the extended Hilbert space.
For simplicity, let us focus on the states \eqref{eq: gaugeinv S3 state} parameterized as follows:
\begin{equation}
\label{eq: restriction on psi}
    b = c = \sqrt{\frac{1}{2}-a^2} ~ , \qquad d = a ~ .
\end{equation}
As will become clear, this set of states is convenient since the resulting density matrices have support on the identity and the single-site operators in the $x$ direction only.\footnote{This choice was inspired by \cite{LinChicago}. We thank Jennifer Lin for providing the slides of her talk to us.} It will be enough to restrict to the states \eqref{eq: gaugeinv S3 state} with \eqref{eq: restriction on psi} to show that the algebraic entanglement entropy proposed in \cite{Lin:2016fqk} does not reproduce entwinement in general.

Reducing the corresponding 8-dimensional density matrix $\rho_{ext}$ on any of the three spins via standard partial tracing yields the following $2\times 2$ reduced density matrix:
\begin{equation}
\rho_{red} = 
    \begin{pmatrix}
    \frac{1}{2} & \frac{1}{3} \left( 1 - 2 a^2 + a \sqrt{6 - 12 a^2}\right)  \\
\frac{1}{3} \left(1 - 2 a^2 + a \sqrt{6 - 12 a^2}\right)   & \frac{1}{2}
    \end{pmatrix} ~ ,
\end{equation}
with eigenvalues
\begin{align}
    \lambda_1 &=  \frac{1}{6} (3 - 2 \sqrt{-(-1 + 2 a^2) (1 + 2 a (2 a + \sqrt{6 - 12 a^2}))}) ~ , \nonumber \\ 
    \lambda_2 &= \frac{1}{6} (3 + 2 \sqrt{-(-1 + 2 a^2) (1 + 2 a (2 a + \sqrt{6 - 12 a^2}))}) ~ .
    \label{eq: eigenvalues reduced density matrix}
\end{align}
The reduced density matrix $\rho_{red}$ has the property that it correctly reproduces the expectation value of (symmetric) single-site operators,
\begin{equation}
\label{eq: check exp values}
    \bra{\psi} \sigma_i^{S} \ket{\psi}={\rm Tr}[\rho^{ext}  \sigma_i^S] \equiv {\rm Tr}[\rho^{ext} \frac{1}{3}\left( \sigma_i^1  + \sigma_i^2 + \sigma_i^3\right)] = {\rm Tr}[\rho_{red} \sigma_i ] ~ ,
\end{equation}
for $i=x,y,z$; and provides a definition for entwinement through its von Neumann entropy,
\begin{equation}
\label{eq: entwinement via extended Hilbert space}
    S_{entw} = -{\rm Tr}[\rho_{red} \log \rho_{red} ] = - \lambda_1 \log \lambda_1 - \lambda_2 \log \lambda_2 ~ ,
\end{equation}
with $\lambda_i$ given by \eqref{eq: eigenvalues reduced density matrix}.

We now turn to the algebraic entanglement entropy approach proposed in \cite{Lin:2016fqk} and examine whether this construction reproduces \eqref{eq: entwinement via extended Hilbert space}. 
As reviewed in section~\ref{subsec:DefinitionsOfEntwinement}, the first step consists in constructing a subalgebra $\mathcal{A}_{\mathcal{O}_{\rho}}$ generated by the projection of $\rho^{S}$ in \eqref{eq: 4d density matrix} on the symmetric single-site operators $\sigma_i^S$ defined in \eqref{eq: check exp values}. (From now on, we consider the matrices $\sigma_i^S$ in their symmetric representation, i.e., as $4 \times 4$ matrices.)
Subsequently, one is instructed to find the unique element $\rho_{red}^S$ of $\mathcal{A}_{\mathcal{O}_{\rho}}$ which reproduces the expectation value of every operator in $\mathcal{A}_{\mathcal{O}_{\rho}}$ in the state $\ket{\psi}$. 
Note that $\mathcal{A}_{\mathcal{O}_{\rho}}$ is a subalgebra of gauge-invariant $4 \times 4$ matrices. 
According to \cite{Lin:2016fqk}, the von Neumann entropy of $\rho_{red}^S$ should equal \eqref{eq: entwinement via extended Hilbert space} with \eqref{eq: eigenvalues reduced density matrix}.  

We start by computing the support of $\rho_{S}$ on the single-site operators
\begin{align}
    p_i \equiv \frac{{\rm Tr}[\rho^{S}  \sigma_i^S]}{{\rm Tr}[\left(  \sigma_i^S\right)^2]} ~ ,
\end{align}
and find
\begin{align}
    p_x &= \frac{3}{10} (1 - 2 a^2 + a \sqrt{6 - 12 a^2}) ~ , \\
    p_y &= p_z = 0 ~ .
\end{align}
Clearly, the simple support of the density matrix follows from our choice \eqref{eq: restriction on psi}.
As a second step, we are instructed to define 
\begin{equation}
\label{eq: Orho}
    \mathcal{O}_\rho = p_x \sigma_x^S
\end{equation}
as the generator of the gauge-invariant subalgebra $\mathcal{A}_{\mathcal{O}_{\rho}}$,
\begin{equation}
    \mathcal{A}_{\mathcal{O}_{\rho}} = \text{span}\{ \mathbf{1}, \mathcal{O}_\rho,\mathcal{O}_\rho^2, \mathcal{O}_\rho^3\} ~ ,
\end{equation}
which we find closes after including the third power of \eqref{eq: Orho}. 
Note that the operators $\{\mathcal{O}_\rho^i\}$ are not mutually orthogonal. Hence, in order to ensure that we find a reduced density matrix 
\begin{equation}
\label{eq: expansion red in generators}
    \rho_{red}^S = \sum_{i=0}^3 c_i \mathcal{O}_\rho^i
\end{equation}
which reproduces the expectation value of any element in $\mathcal{A}_{\mathcal{O}}$ in the state $\ket{\psi}$, we introduce an inner product
\begin{equation}
    M_{ij} = {\rm Tr}[\mathcal{O}_\rho^{i}\mathcal{O}_\rho^j] ~ .
\end{equation}
Then, one can obtain the coefficients of \eqref{eq: expansion red in generators} as
\begin{equation}
\label{eq: coefs of rhored}
    c_i  = \sum_{j=0}^3 M^{-1}_{ij} {\rm Tr}\left(\mathcal{O}_\rho^{j} \rho^S \right) ~ .
\end{equation}

Finally, the claim of \cite{Lin:2016fqk} is that the von Neumann entropy of the reduced density matrix \eqref{eq: expansion red in generators} with \eqref{eq: coefs of rhored} reproduces \eqref{eq: entwinement via extended Hilbert space} with \eqref{eq: eigenvalues reduced density matrix}. We find that $\rho_{red}^S$ has two vanishing and two nonzero eigenvalues:
\begin{align}
    \tilde{\lambda}_1 &= \frac{1}{4} +  a^2 - a\sqrt{\frac{3 - 6 a^2}{2}} ~ ,
    & \tilde{\lambda}_2 &= \frac{3}{4} -  a^2 + a\sqrt{\frac{3 - 6 a^2}{2}} ~ .
\end{align}
As a result, the corresponding von Neumann entropy
\begin{equation}
\label{eq: algebraic entanglement entropy}
    S_{AEE} = -{\rm Tr}[\rho^S_{red} \log \rho^S_{red} ] = - \tilde{\lambda}_1 \log \tilde{\lambda}_1 - \tilde{\lambda}_2 \log \tilde{\lambda}_2 ~ ,
\end{equation}
which is an algebraic entanglement entropy in the usual sense, differs from \eqref{eq: entwinement via extended Hilbert space} with \eqref{eq: eigenvalues reduced density matrix}. This provides a counterexample to the generality of the construction in \cite{Lin:2016fqk}.

\bibliographystyle{apsrev4-2}
\bibliography{references.bib}

\end{document}